# Improved LM Test for Robust Model Specification Searches in Covariance Structure Analysis: Application in Political Science Research

Bang Quan Zheng†
University of Texas at Austin

Peter M. Bentler‡
UCLA

2024

Abstract

Covariance Structure Analysis (CSA) or Structural Equation Modeling (SEM) is critical for political scientists measuring latent structural relationships, allowing for the simultaneous assessment of both latent and observed variables, alongside measurement error. Well-specified models are essential for theoretical support, balancing simplicity with optimal model fit. However, current approaches to improving model specification searches remain limited, making it challenging to capture all meaningful parameters and leaving models vulnerable to chance-based specification risks. To address this, we propose an improved Lagrange Multipliers (LM) test incorporating stepwise bootstrapping in LM and Wald tests to detect omitted parameters. Monte Carlo simulations and empirical applications underscore its effectiveness, particularly in small samples and models with high degrees of freedom, thereby enhancing statistical fit.

[Word Count: 8,834]

---

† Corresponding author. Email: bang.zheng@austin.utexas.edu. ORCID: 0000-0003-2614-2501. Annette Strauss Institute for Civic Life, Moody College of Communication, University of Texas at Austin. 2504 A Whitis Avenue (R2000), Austin, TX 78712.

‡ Departments of Psychology & Statistics, UCLA, ORCID: 0000-0002-9440-721X

## 1 Introduction

Political scientists often grapple with complicated and abstract concepts such as democracy, value, ideology, identity, trust, and political tolerance, among others (Goren 2005; Sullivan et al. 1981; Davidov 2009; Acock, Clarke, and Stewart 1985; Pietryka and MacIntosh 2013; Feldman 1988). They may utilize Covariance Structure Analysis (CSA) or Structural Equation Modeling (SEM) with latent variables, such as confirmatory factor analysis, to estimate statistical models that amalgamate indicators, aiming to gauge the underlying latent concepts. SEM's appeal lies in its dual capability to assess multiple hypotheses regarding the influences of latent and observable variables on other variables, while also enabling simultaneous modeling of measurement error (Yuan and Liu 2021; Blackwell, Honaker, and King 2017). For instance, SEM has been successfully employed in studying diverse relationships, including the connection between party identification and core values (Goren 2005), political conceptualization (Zheng 2023), political tolerance and democracy theory (Sullivan et al. 1981), values and support for immigration (Davidov 2009), the underlying dimensions of racial attitudes (DeSante and Smith 2022), and measurement invariance analysis (Davidov 2009; Oberski 2014; Pietryka and MacIntosh 2013).

Due to its versatility, CSA or SEM has shown a modest yet consistent trend in usage within political science research over the past decades. Figure 1 shows the frequency of articles involving SEM across six political science journals. Data collection, conducted through Google Scholar advanced search, spans from 1990 to 2020. The keywords used were "structural equation modeling," "covariance structure analysis," and "factor analysis." Figure 1 illustrates that between 1990 and 2020, the number of articles utilizing SEM increased in a nearly linear fashion, starting at approximately 10 articles per year and rising to around 20 articles per year. This trend underscores a sustained interest in applying this methodology, particularly within political psychology.

Figure 1. Number of Articles Published in Selected PS Journals Using SEM

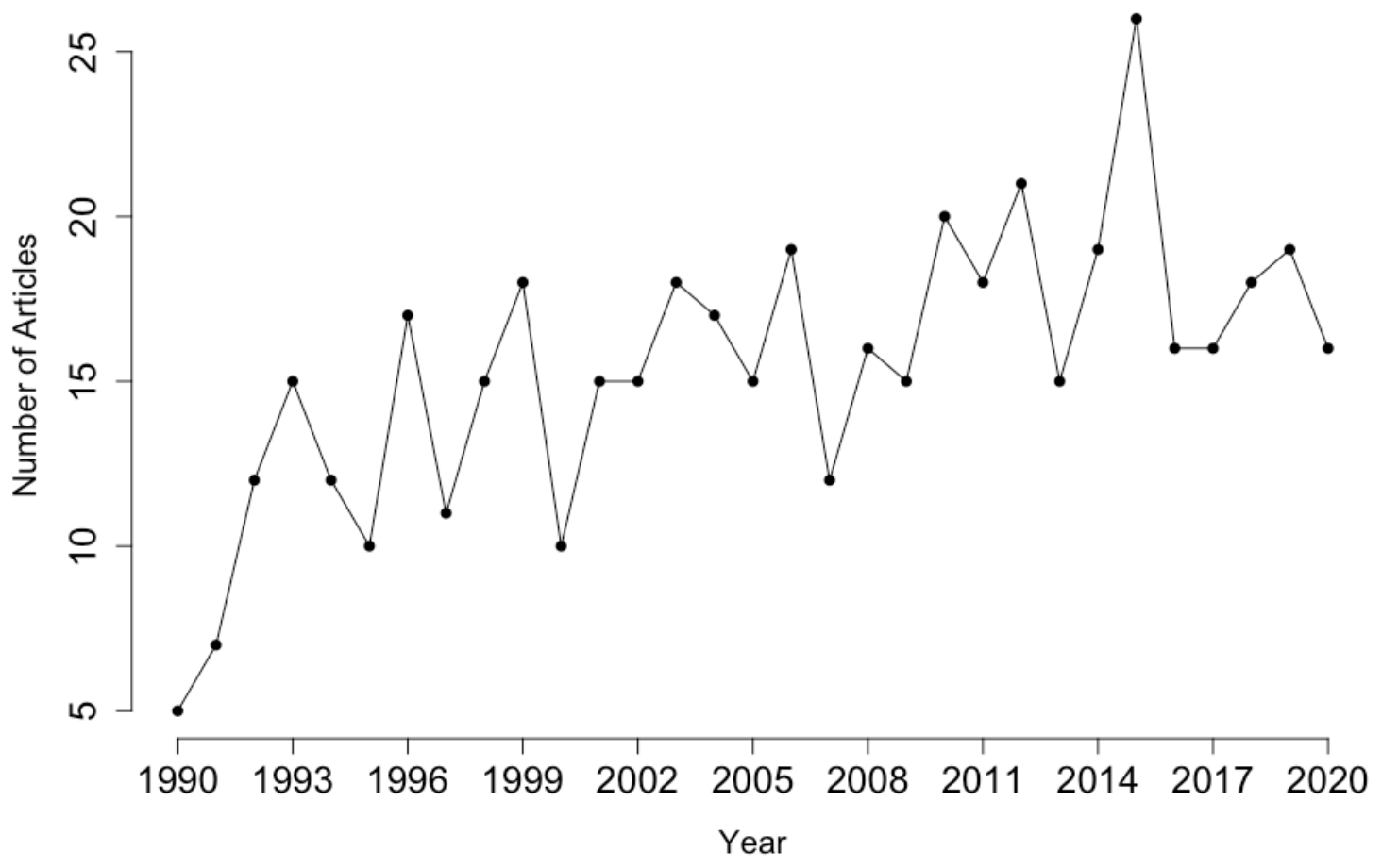

Note: The data are based on a Google Scholar advanced search covering the years 1990 to 2020, focusing on publications in *The American Political Science Review*, *American Journal of Political Science*, *The Journal of Politics*, *Political Psychology*, *Political Behavior*, and *Public Opinion Quarterly*.

As with any modeling technique, the adequate specification of CSA or SEM models is critical for making sound decisions and drawing valid inferences (Zheng and Bentler 2024). Identifying suitable parameters to fit complex models becomes particularly challenging when dealing with a large number of observed variables relative to small sample sizes. In such instances, there's a heightened chance-based model risks (a.k.a. capitalizing on chance in psychometric literature), which can compromise the reliability of the findings (Bentler 2006; Yuan and Liu 2021; Sörbom 1989). This pervasive issue highlights the necessity for robust techniques that can enhance the stability of these models. Unlike regression models, where the focus is primarily on the relationship between the dependent and key independent variables, with coefficients holding the most weight, SEM can handle intricate models with numerous interrelated variables and pathways. It facilitates the examination of complex theoretical frameworks. Therefore, researchers' arguments rely on the underlying structural relationships, rendering both model specification and fit equally crucial.

This study proposes a novel method, the improved Lagrange Multipliers (LM) test, for model specification searches, addressing the challenge of noise interference. Our data-driven specification search method, using the stepwise bootstrap approach in both LM and Wald tests, effectively identifies potential omitted parameters, improving the precision of parameter identification. Through a series of simulation studies and two empirical applications in political science, our results demonstrate that the improved LM test is particularly reliable when dealing with small sample sizes in models with high degrees of freedom. The improved LM test enhances the reliability, validity, and statistical fit of model specifications while mitigating the risk of being misled by noise, enabling researchers to draw sound conclusions grounded in solid statistical evidence.

As we will demonstrate later with empirical examples from Huddy and Khatib (2007), Davidov (2009), and Oberski (2014), the theoretical arguments in these studies are grounded in structural relationships among sets of latent and observed variables. Inadequate model specification could undermine these arguments, whereas a well-specified model with robust goodness-of-fit can reinforce them. For instance, Huddy and Khatib argue that national identity is distinct from other forms of national attachment, such as symbolic, constructive, uncritical patriotism, and nationalism, and that a strong American identity promotes civic involvement. However, the weak $\chi^2$ test statistic reported in their research may call this claim into question. Moreover, the new parameter identified through the improved LM test not only reinforces Huddy and Khatib's original argument but also underscores the significant role of national identity in driving emotional reactions, such as anger—an aspect that was overlooked in their original model. In Davidov's (2009) and Oberski's (2014) models, we focused exclusively on the German sample and identified two omitted variables that indicate potential model misspecification. While this misspecification may not be substantial enough to alter the substantive conclusions, including these variables strengthens the authors' arguments by improving model fit. If untested, confounding measurement inequivalence with structural differences could lead to specification issues.

## 2 Challenges and Existing Approaches in Model Specification Searches

In this section, we review existing approaches for model specification searches, covering major tests, their procedures, and model fit evaluations. To evaluate the model fit between the theoretical model and sample data, researchers must assess the model's adequacy using goodness-of-fit tests. However, before trusting the $\chi^2$ test statistics and other fit indices, adequate model modification and specification are necessary. In CSA or SEM, a desirable model fit involves striking a balance between simplifying the model without compromising the overall fit and improving the model fit without making it more complicated (Bentler and Chou 1992; MacCallum, Roznowski, and Nectowitz 1992). Several critical factors can affect overall model fit, such as poorly specified models. Typically, researchers specify a model based on priori knowledge and fit it to sample data by estimating parameters. To modify the model, researchers determine the number of parameters to add or remove from the existing model and then refit it with the same dataset. If the initial model fit is inadequate, a common practice is to free parameter restrictions to enhance the model's fit to the data (Bentler and Chou 1992; Kaplan 1988; Leamer 1978; MacCallum, Roznowski, and Nectowitz 1992; Sörbom 1989). This process is referred to as model specification search.

The goal of model specification searches and modifications is to develop a generalizable model that demonstrates stability. Stability refers to the consistency of model results across repeated samples (MacCallum, Roznowski, and Nectowitz 1992). While it is challenging to achieve a perfect model in practice, an acceptable model specification should consist of a set of parameters supported by substantive theories that also has an adequate statistical fit (Bentler and Chou 1992; Chou and Huh 2012; MacCallum, Roznowski, and Nectowitz 1992; Yuan, Hayashi, and Yanagihara 2007).

The process of modifying and specifying any statistical models can be influenced by idiosyncratic characteristics of the data, meaning that modifications and specifications that improve the fit of one model may not necessarily apply to another random sample from the same population. This challenge, often referred to as the chance-based model risks, becomes particularly pronounced in large models with high degrees of freedom but relatively small sample sizes (MacCallum, Roznowski, and Nectowitz 1992). In such cases, increased sampling variability in the sample covariance can significantly impact the results of CSA or SEM analyses, leading to inconsistencies across different samples. Despite the significance of this issue, there are currently no systematic approaches to enhance model modification and specification in CSA or SEM, making it challenging to include all statistically meaningful parameters in the model without the interference of noise.

Two key considerations for assessing model adequacy are model parsimony and model fit. Model parsimony refers to the number of free parameters in the model, while model fit is evaluated using empirical fit indices. Poor model fit can occur in two scenarios: if a model inadequately fits the data (under-specified), requiring modifications by releasing constraints on fixed parameters in a "forward search," or if a model fits the data well but has excessive parameters (overfitting), necessitating simplification through constraints on free parameters in a "backward search."

## *2.1 The LM Test*

In multivariate analysis of CSA or SEM, two commonly used test statistics are the LM test and the Wald test. The LM test only requires estimating the restricted model, while the Wald test requires a more comprehensive model. Notably, the statistical theory for the LM test is more complex than for the Wald test. This study focuses on estimating the restricted model under various constraints. The LM test is particularly useful for guiding model modifications to improve fit, as it identifies the effects of freeing initially fixed parameters (Lee and Bentler 1980; Bentler 1986; Sörbom 1989; Satorra 1989; Yuan and Liu 2021).

Standard LM tests rely on a single snapshot of the initial model, which may not accurately identify missing parameters in population data. Consequently, model misspecifications can occur, leading to poor generalization to new samples. This limitation is particularly evident with small sample sizes, as the effectiveness of model modification using the LM test becomes compromised and susceptible to random variations (MacCallum, Roznowski, and Nectowitz 1992; Yuan and Liu 2021).

A model is deemed acceptable when its parameters align with theories and show good statistical fit to the data (Chou and Huh 2012). If a model has $q$ free parameters and an additional nondependent variable $r$, where $r < q$, the LM test can be used to identify which fixed parameters should be freed for better model fit. This is done by computing the LM test statistic $T_{LM}$. The LM test employs forward specification searching, where a constraint in the initial model is proposed to be freed based on how much it would enhance model fit.

A model consists of both free and fixed parameters, with the latter included to specify the model. Let $\hat{\boldsymbol{\theta}}$ be a vector of constrained estimators of $\boldsymbol{\theta}$ that satisfies the $r < q$ constraints $h(\boldsymbol{\theta})$=0 when minimizing the fit function $F(\boldsymbol{\theta})$ for a given model. This is equivalent to minimizing the function of a constrained model while assuming $h(\boldsymbol{\theta}) = \theta_r = 0$. With $r$ constraints, there exists an $r \times 1$ constraint vector $h(\boldsymbol{\theta})' = (h_1, \cdots, h_r)$. When minimizing the fit function $F(\boldsymbol{\theta})$ with constraints of $h(\boldsymbol{\theta})$=0, matrices of derivatives $\boldsymbol{g} = (\frac{\partial F}{\partial \boldsymbol{\theta}})$ and $\boldsymbol{L}' = (\frac{\partial h}{\partial \boldsymbol{\theta}})$ exist for the "forward search," and there will be a vector of LM multipliers, λ, such that

$$\hat{\boldsymbol{g}} + \hat{\boldsymbol{L}}'\hat{\boldsymbol{\lambda}} = 0 \text{ and } h(\hat{\boldsymbol{\theta}}) = 0. \quad (1)$$

For the LM test to be applicable, several technical regularity conditions must be met, including the continuity of $\partial h/\partial \boldsymbol{\theta}$, model identification, positive definiteness of $\boldsymbol{\Sigma}$, linearly independent constraints, and full rank matrices of derivatives $\boldsymbol{g}$ and $\boldsymbol{L}'$. In the context of constraints, the asymptotic covariance matrix can be derived from an information matrix $\boldsymbol{H}$, augmented by the matrix of derivatives $\boldsymbol{L}$ and a null matrix **O**. Thus, the sample variance covariance matrix of the estimated parameter, $\sqrt{n}(\hat{\boldsymbol{\theta}} - \boldsymbol{\theta})$, is given by the inverse of the Fisher information matrix of $\boldsymbol{H}(\boldsymbol{\theta})$, associated with $q$ free parameters in $\boldsymbol{\theta}$ in the case of maximum likelihood estimation, and $\boldsymbol{R}$ gives the covariance matrix of the Lagrange multipliers $\sqrt{n}(\hat{\boldsymbol{\lambda}} - \boldsymbol{\lambda})$, which is derived from the inverse of the information matrix $\boldsymbol{L}$. Therefore, we can define

$$\begin{bmatrix} \boldsymbol{H} & \boldsymbol{L}' \\ \boldsymbol{L} & \boldsymbol{O} \end{bmatrix}^{-1} = \begin{bmatrix} \boldsymbol{H}^{-1} - \boldsymbol{H}^{-1}\boldsymbol{L}'(\boldsymbol{L}\boldsymbol{H}^{-1}\boldsymbol{L}')^{-1} & \boldsymbol{H}^{-1}\boldsymbol{L}'(\boldsymbol{L}\boldsymbol{H}^{-1}\boldsymbol{L}')^{-1} \\ (\boldsymbol{L}\boldsymbol{H}^{-1}\boldsymbol{L}')^{-1}\boldsymbol{L}\boldsymbol{H}^{-1} & -(\boldsymbol{L}\boldsymbol{H}^{-1}\boldsymbol{L}')^{-1} \end{bmatrix} = \begin{bmatrix} \boldsymbol{M} & \boldsymbol{T}' \\ \boldsymbol{T} & -\boldsymbol{R} \end{bmatrix}. \quad (2)$$

Under regularity conditions and the null hypothesis $\boldsymbol{H}(\boldsymbol{\theta})$, the LM test is available in two versions, and the multivariate LM statistics are asymptotically distributed as a $\chi^2$ variate with $r$ degrees of freedom. They compute all constraints simultaneously.

$$T_{LM} = n\hat{\boldsymbol{\lambda}}'\hat{\boldsymbol{R}}^{-1}\hat{\boldsymbol{\lambda}} \sim \chi^2_r. \quad (3)$$

A univariate LM statistic is used to test a single constraint and is distributed as a $\chi^2$ variate with 1 $df$. This test is particularly useful for evaluating whether a specific parameter in the $\boldsymbol{\theta}_r$ vector is equal to 0:

$$T_{LMi} = n\hat{\boldsymbol{\lambda}}_i^2\hat{\boldsymbol{R}}^{-1}\hat{\boldsymbol{\lambda}}_i \sim \chi^2_{r1}. \quad (4)$$

This is also known as a modification index in the LISREL program (Jöreskog and Sörbom 1988). For Equations 3 and 4, $r$ is the number of nondependent constraints, and the matrices $\boldsymbol{H}$ and $\boldsymbol{L}'$ have dimensions $q \times q$ and $q \times r$, respectively.

$$\hat{\boldsymbol{\lambda}} = \left(\hat{\boldsymbol{L}}\hat{\boldsymbol{H}}^{-1}\hat{\boldsymbol{L}}'\right)^{-1}\hat{\boldsymbol{L}}\hat{\boldsymbol{H}}^{-1}\hat{\boldsymbol{g}}. \quad (5)$$

The LM test performance depends on factors like sample size, degrees of freedom, and the number of variables and parameters. Degrees of freedom are influenced by the number of parameters and variables. Higher $p$ and lower $q$ result in more degrees of freedom. The LM test considers all possible paths based on the degrees of freedom. With more degrees of freedom, the number of possible paths increases, increasing the risk of falsely identifying non-existent paths in the true model, especially when there are few missing paths. However, this risk decreases with larger sample sizes, reducing reliance on chance occurrences.

## 2.2 *The Wald Test*

The Wald test follows a backward stepwise procedure to identify which free parameter, starting with the one with the smallest Wald test statistic, should be removed from the model. This is done by including candidate parameters and performing univariate Wald tests on each. Parameters with statistically significant Wald test values are retained, and the process continues until no further parameters can be added. The Wald test statistic is calculated as follows:

$$W = n\widehat{\boldsymbol{\theta}}_{\boldsymbol{r}}'(\widehat{\boldsymbol{L}}\widehat{\boldsymbol{H}}^{-1}\widehat{\boldsymbol{L}}')\widehat{\boldsymbol{\theta}}_{\boldsymbol{r}} \sim \chi_r^2, \tag{6}$$

where $\widehat{\boldsymbol{L}}$ is a quadratic form. The closer $\widehat{\boldsymbol{L}}$ is to 0, the more likely it is that the null hypothesis equals $0$ will be rejected. The univariate Wald statistics $\hat{\theta}_i$ for each of the parameters in $\widehat{\boldsymbol{\theta}}_{\boldsymbol{r}}$ can be expressed as

$$W_i = n\hat{\theta}_i\widehat{H}_{ii}^{-1},\ \hat{\theta}_i = n\theta_i^2/\widehat{H}_{ii} \sim \chi_i^2\ , \tag{7}$$

where $\hat{\theta}_i$ is one of the parameters in $\widehat{\boldsymbol{\theta}}_{\boldsymbol{r}}$ and $\widehat{H}_{ii}$ is the *i*th parameter in the diagonal of the $\boldsymbol{H}$ matrix. The computational complexity of the LM and Wald tests depends on the number of free and fixed parameters in a model. As the number of parameters increases, estimating and comparing their effects becomes more computationally demanding. The $\boldsymbol{H}$ matrix, which represents the covariance matrix of the independent variables, reflects the number of parameters that can be either free or fixed. For a model with *k* independent variables, the $\boldsymbol{H}$ matrix will have $k^2$ elements (Chou and Huh 2012).

*2.3 Evaluation of Model Fit*

This study uses maximum likelihood (ML) estimation, the standard method for deriving goodness-of-fit statistics and parameter estimates in CSA under normal theory. In CSA, a random sample, $x \in \{x_1, \dots., x_n\}$ is assumed to be independently and identically distributed, following a multivariate normal distribution $\mathcal{N}[\boldsymbol{\mu}, \boldsymbol{\Sigma}_0]$. Here, $\boldsymbol{\mu}$ represents a vector of sample means, and the covariance matrix $\boldsymbol{\Sigma}_0$ is assumed positive definite with an unknown population parameter vector $\boldsymbol{\theta}_0$ of dimension $q \times 1$, where $\boldsymbol{\Sigma}_0 = \boldsymbol{\Sigma}(\boldsymbol{\theta}_0)$. The sample covariance matrix is:

$$\boldsymbol{S} = \frac{1}{n-1} \sum_{i=1}^{n} (x_i - \bar{x})(x_i - \bar{x})' \tag{8}$$

where the sample mean $\bar{x} = \frac{1}{n} \sum_{i=1}^{n} ( x_1, \dots., x_n)$. $\boldsymbol{S}$ serves as an unbiased estimator of the population covariance $\boldsymbol{\Sigma}_0$.

In CFA, a model can be represented as:

$$\boldsymbol{x_i} = \boldsymbol{\mu} + \boldsymbol{\Lambda \xi_i} + \boldsymbol{\epsilon_i}, \qquad i = 1, \dots, n \tag{9}$$

where $\boldsymbol{x_i}$ is a random sample, $\boldsymbol{\mu}$ is a sample mean vector, $\boldsymbol{\Lambda}$ is a matrix of factor loadings, $\boldsymbol{\xi_i}$ is a vector of latent factors, and $\boldsymbol{\epsilon_i}$ is a vector of residuals. Here, the parameters involved in a model are contained in the covariance matrix $\boldsymbol{\Sigma}$ of the observed variables. $\boldsymbol{\Sigma} = \boldsymbol{\Lambda \Phi \Lambda}' + \boldsymbol{\Psi}$, where $\boldsymbol{\Lambda}$ again is a factor loading matrix, and $\boldsymbol{\Phi}$ is a covariance matrix of the latent factors, and $\boldsymbol{\Psi}$ is a covariance matrix of unique scores.

The population covariance matrix $\boldsymbol{\Sigma}$ is modeled as $\boldsymbol{\Sigma}(\boldsymbol{\theta})$, where $\boldsymbol{\theta}$ contains free parameters $\boldsymbol{\Lambda}$, $\boldsymbol{\Phi}$, and $\boldsymbol{\Psi}$. The sample covariance matrix $\boldsymbol{S}$ serves as an unbiased estimator of $\boldsymbol{\Sigma}$, with the null hypothesis $\boldsymbol{\Sigma} = \boldsymbol{\Sigma}(\boldsymbol{\theta})$. An objective function $F[\boldsymbol{\Sigma}(\boldsymbol{\theta}), \boldsymbol{S}]$ measures the discrepancy between $\boldsymbol{\Sigma}(\boldsymbol{\theta})$ and $\boldsymbol{S}$.

This study derives the goodness-of-fit statistic $T_{ML}$ using the ML discrepancy function (Jöreskog 1969), fitting the model-implied covariance matrix $\boldsymbol{\Sigma}(\boldsymbol{\theta})$ to the sample covariance matrix $\boldsymbol{S}$, as shown in Equation 10:

$$F_{ML}(\theta) = \log|\boldsymbol{\Sigma}\ (\boldsymbol{\theta}) - \log|\boldsymbol{S}| + tr(\boldsymbol{S}\boldsymbol{\Sigma}(\boldsymbol{\theta})^{-1}) - p \tag{10}$$

$$\widehat{\boldsymbol{\theta}}_{ML} = argmin\ F_{ML}(\boldsymbol{\theta}) \tag{11}$$

The ML fit function $F_{ML}(\boldsymbol{\theta})$ derives parameter estimates in $\boldsymbol{\Sigma}(\boldsymbol{\theta})$ that minimize the test statistic. At its minimum, as shown in Equation 11, $\widehat{\boldsymbol{\theta}}_{ML}$contains parameter estimates $\widehat{\boldsymbol{\Lambda}}$, $\widehat{\boldsymbol{\Phi}}$, and $\widehat{\boldsymbol{\Psi}}$. Using these, we can reconstruct the sample covariance matrix to align with the model-implied covariance $\boldsymbol{\Sigma}(\widehat{\boldsymbol{\theta}}) = \widehat{\boldsymbol{\Lambda}}\widehat{\boldsymbol{\Phi}}\widehat{\boldsymbol{\Lambda}}' + \widehat{\boldsymbol{\Psi}}$, assuming the sample-implied matrix matches the population matrix. If $\boldsymbol{S} \approx \boldsymbol{\Sigma}(\widehat{\boldsymbol{\theta}})$ with *p*-value > 0.05, the model is considered plausible.

The ML goodness-of-fit test statistic is calculated as:

$$T_{ML} = (N-1)F_{ML}(\widehat{\boldsymbol{\theta}}). \tag{12}$$

This statistic is the product of $F_{ML}(\widehat{\boldsymbol{\theta}})$ and $(N-1)$, where $N$ is sample size. As $N$ increases, $T_{ML}$ is expected to asymptotically follow a $\chi^2$ distribution with corresponding degrees of freedom.

## 3 Improved LM Test

To overcome the limitations in existing model specification searches, we propose a novel approach leveraging bootstrap methods for data-driven model specification searches, integrating the LM and Wald tests. It involves generating multivariate random samples through bootstrap resampling based on the initial model. We start with a forward stepwise bootstrap resampling method in the standard LM test. Following this, using the statistically significant results from the bootstrap LM, we apply a backward stepwise bootstrap Wald test to mitigate overfitting by identifying potential paths that may not be needed. This iterative workflow strikes a balance between maximizing model fit, which the LM test emphasizes, and maintaining parsimony, as the Wald test tends to emphasize. We term this approach the "improved LM test," offering a valuable tool for enhancing model fit and reducing chance-based model risks in applied CSA or SEM.

The improved LM approach for specification searches involves a multi-stage process. Initially, we create a hypothetical confirmatory factor analysis (CFA) population model and generate multivariate normal data with varying sample sizes using the Monte Carlo method. We then construct a misspecified analysis model by omitting several true parameters from the population model and fit it to the simulated data. To identify missing parameters in the measurement model, we conduct a univariate LM test to detect potential omissions. The parameters are ranked by their $\chi^2$ statistics, and we select a series of them as the testing parameters. Next, we perform a multivariate stepwise LM test. This forward stepwise procedure follows a *general-to-specific* approach in specification searches within spatial econometrics (Mur and Angulo 2009; Florax, Folmer, and Rey 2003). To enhance the reliability of LM test results, each sequence in the forward stepwise LM procedure is based on bootstrap resampling of the initial data. We calculate the means of the bootstrap LM test statistics and *p*-values, selecting parameters with *p*-values $< 0.05$ for further testing using bootstrap Wald tests to assess their stability.

In the Wald test procedure, we conduct a backward stepwise search by initially including all LM-based parameters in the model and then sequentially fixing one parameter at a time. This *specific-to-general* approach employs bootstrap resampling to compute the mean $\chi^2$ test statistics and *p*-values. However, applying the Wald test to all missing parameters is generally impractical for midsize to large models, as the number of possible omitted parameters may exceed the sample size, resulting in a singular matrix that cannot be inverted. Limiting the Wald test to bootstrap LM-based selected parameters resolves this issue. We include the bootstrap LM-based parameters in the model and perform univariate Wald tests on each. By focusing on these parameters, we significantly reduce the number of elements in the ***H*** matrix in Equation (6) related to variances and covariances, as compared to a full model with all potential missing parameters. This reduction in ***H*** matrix elements facilitates efficient matrix inversion, leading to faster computations and more stable estimates. Thus, unlike other specification search methods in spatial econometrics, the improved LM test integrates LM and Wald tests with bootstrap resampling to reduce noise, enhance reliability, and distinguish between meaningful results and random outcomes. A detailed illustration is provided in the next section.

## 4 Simulation & Multi-Stage Process for Model Specification Searches

In this section, we illustrate the improved LM test for model specification, describing the setup of hypothetical population and analysis models, the simulation procedure, and a multi-stage process for conducting model specification searches that integrates bootstrap sampling within the LM and Wald tests.

### *4.1 Population Model and Analysis Model*

The simulation begins with a hypothetical population model consisting of a three-factor structure, with each factor measured by eight manifest variables, as illustrated in Figure 2. An analysis model, shown in Figure 3, consists of three factors, each linked to eight indicators, with all factors freely correlated, resulting in a total of 24 variables. The four dashed lines in Figure 2 represent the paths not included in the analysis model. Omitting four parameters aims to reduce the chance of the initial model closely resembling the true model. Including many missing parameters can result in a poorly specified model, potentially rendering LM test results meaningless or falsely indicating statistical significance by chance (Yuan, Marshall, and Bentler 2003).

Figure 2. Path Diagram of the Population Model

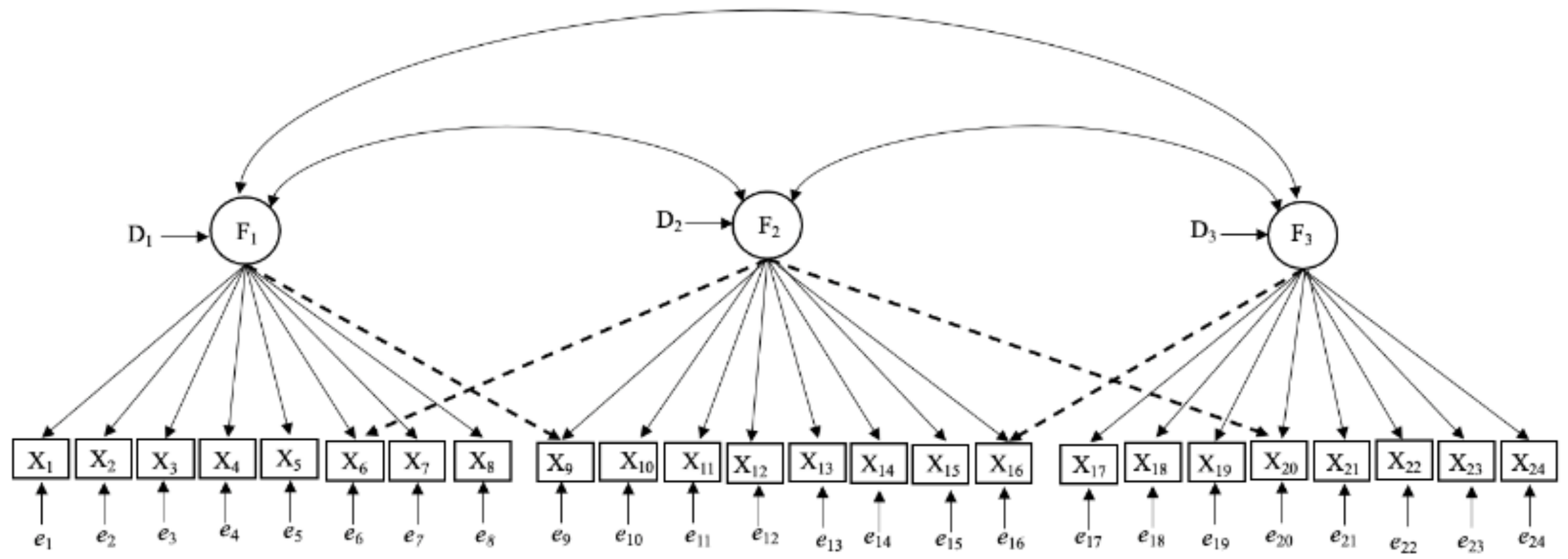

Figure 3. Path Diagram of the Misspecified Analysis Model

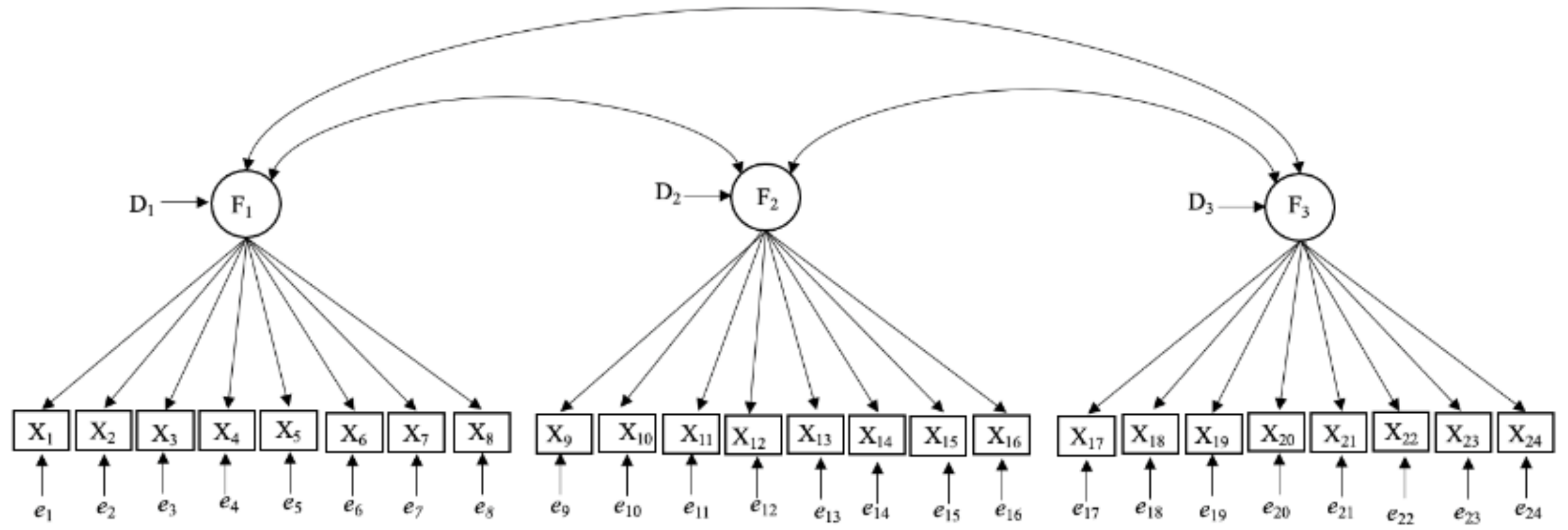

## *4.2 Monte Carlo Simulations*

The simulated data are generated using a standard confirmatory factor model, given by Equation (13):

$$\boldsymbol{x_i} = \boldsymbol{\Lambda\xi_i} + \boldsymbol{\epsilon_i} \tag{13}$$

where $\boldsymbol{x_i} = (x_{i1}, x_{i2}, \dots x_{ip})'$ is a vector of $p$ observations on person $i$ in a population, and $i = 1,2, \dots n$. $\boldsymbol{\Lambda}$ is a matrix of factor loadings, and $\boldsymbol{\epsilon_i} = (\epsilon_{i1,}\epsilon_{i2,} \dots . \epsilon_{ip})'$ is a vector of error terms, and $\text{var}(\epsilon) = \boldsymbol{\Psi}$. $\boldsymbol{\xi_i} = (\xi_{i1,}\xi_{i2,\dots}\xi_{im})'$ is a vector of latent factors, and $\text{var}(\boldsymbol{\xi}) = \boldsymbol{\Phi}$. Each latent factor $\boldsymbol{\xi_i}$ has a mean and a variance and may correlate with other latent factors $\boldsymbol{\xi_i}$; whereas $\boldsymbol{\xi_i}$ and $\boldsymbol{\epsilon_i}$ are uncorrelated, so that $E(\boldsymbol{\xi}) = \boldsymbol{\mu}_{\xi}$, which is the mean of the factors.

With the data generation scheme and population model described above, we simulate a population and draw samples using Monte Carlo simulation, based on the predefined matrices $\boldsymbol{\Lambda}'$ and $\boldsymbol{\Phi}$:

$$\boldsymbol{\Lambda}' = \begin{bmatrix} 0.65 & 0.65 & 0.7 & 0.7 & 0.7 & 0.7 & 0.6 & 0.5 & 0.5 & 0 & 0 & 0 & 0 & 0 & 0 & 0 & 0 & 0 & 0 & 0 & 0 & 0 & 0 & 0 \\ 0 & 0 & 0 & 0 & 0 & 0.5 & 0 & 0 & 0.6 & 0.6 & 0.6 & 0.7 & 0.7 & 0.5 & 0.5 & 0.65 & 0 & 0 & 0 & 0.65 & 0 & 0 & 0 & 0 \\ 0 & 0 & 0 & 0 & 0 & 0 & 0 & 0 & 0 & 0 & 0 & 0 & 0 & 0 & 0 & 0.45 & 0.5 & 0.5 & 0.5 & 0.6 & 0.6 & 0.6 & 0.7 & 0.7 \end{bmatrix},$$

$$\boldsymbol{\Phi} = \begin{bmatrix} 1 & & \\ 0.3 & 1 & \\ 0.4 & 0.5 & 1 \end{bmatrix}$$

When diag($\boldsymbol{\Sigma}$)=$\mathbf{I}$, which is an identity matrix, the unique variances can be determined by $\boldsymbol{\Psi} = \mathbf{I}_{24} - \text{diag}(\boldsymbol{\Lambda\Phi\Lambda}')$. Since we are not interested in the mean structure, we set the factor means $\mu's = (0, 0, 0)$. The data generating process consists of two steps. 1) We draw from a multivariate normal distribution with zero mean and covariance matrix $\boldsymbol{\Phi}$. Unique factors $\boldsymbol{\epsilon_i}$ are drawn from a multivariate normal distribution with zero mean and covariance $\boldsymbol{\Psi}$. Utilizing Equation (13), this procedure generates multivariate normal observations characterized by a covariance matrix $\boldsymbol{\Sigma}(\boldsymbol{\theta})$.

The data generation and all analyses for this research are conducted using the 'lavaan' package (Version 4.2.3.) (Rosseel 2012) in R, based on the previously specified population and assuming multivariate normality. The simulation studies involved sample sizes of N=100 to 10,000. Our testing models consisted of 24 observed variables (*p*=24) and 3 latent factors, resulting in a covariance component of *p**=24(24+1)/2=300, with 55 free parameters to estimate, and 245 degrees of freedom. This model size is a good representation of most SEM research.

To assess the stability of the model specifications, we conduct a series of Monte Carlo simulations using the population model (Figure 2) across different sample sizes and fit the analysis model (Figure

3). The study includes 12 sample sizes—100, 150, 200, 250, 300, 350, 400, 500, 1,000, and 2,000—that are selected to reveal important phenomena related to the issues under study. To evaluate the goodness-of-fit, we employ various methods, including $\chi^2$ test statistics, standard deviations of the $\chi^2$ test, and the rejection rate. Additionally, we utilize alternative fit measures such as the comparative fit index (CFI), normed fit index (NFI), and root mean square error of approximation (RMSEA). However, as the analysis model does not fit the population model, we expect the $\chi^2$ test statistic to be larger than the degrees of freedom and the $p$-value < 0.05.

### *4.3 Selection of Testing Parameters*

To commence the specification search, an initial set of parameters is required. The process of selecting these parameters begins with an exploratory univariate LM test via model modification indices (Sörbom 1989). However, as the data are generated from Monte Carlo simulations, each sample drawn from the population model will be different, leading to variability in $T_{LMi}$. For instance, if we draw 500 samples of the same sample size, we will obtain 500 unique sets of initial testing parameters. Nonetheless, there should be a set of common parameters that frequently appear across all samples, including the true missing parameters. To obtain a more representative set of initial parameters, after we simulate the data for 500 trials, we calculate the mean $T_{LMi}$ of each parameter and sort them in descending order. We then select the top 12 parameters with the largest $T_{LMi}$ to test the proposed improved LM test. The LM test is designed to enhance the fit of the existing model, assuming it is reasonably well-fitted. Consequently, a sensible model should expect only a few significant omitted parameters. If the count exceeds 12, the model may suffer from severe misspecification issues, necessitating a new formulation.

### *4.4 Bootstrap Simulation*

The bootstrap method efficiently approximates population covariance structures in simulations, providing a practical alternative when the distribution of the sample is unknown. This approach tackles numerous challenges that conventional statistical methods encounter. For instance, the bootstrap approach does not assume normally distributed data. Even in cases where the data are normally distributed, at a given sample size, the bootstrap often provides more accurate results than those based on standard asymptotic methods (Yuan et al., 2007).

Let $x_1, x_2, \cdots, x_n$ denote a sample with a covariance matrix represented as $\boldsymbol{S}$ where its population counterpart is $\boldsymbol{\Sigma_0}$. The bootstrap method iteratively draws samples from a known empirical distribution function, effectively substituting it for the population in the bootstrap samples. However, since the population covariance matrix $\boldsymbol{\Sigma_0}$ is unknown, an alternative matrix $\dot{\boldsymbol{S}}$ must be found to serve as a surrogate for $\boldsymbol{\Sigma_0}$. Consequently, each $x_i$ can be transformed into $x_i'$ by:

$$x_i' = \dot{\boldsymbol{S}}^{1/2}\boldsymbol{S}^{-1/2}x_i,\ i = 1, 2, \cdots, n \tag{14}$$

where $\dot{\boldsymbol{S}}^{1/2}$ $\dot{\boldsymbol{S}}^{1/2}$ is a $p \times p$ matrix satisfying $(\boldsymbol{S}^{-1/2})(\boldsymbol{S}^{-1/2})' = \dot{\boldsymbol{S}}$. The subsequent steps involve generating the bootstrap samples by sampling with replacement from $(x_1', x_2', \cdots, x_n')$, thereby computing the sample covariance matrix denoted as $\boldsymbol{S}^*$ for these bootstrap samples.

### *4.5 Bootstrap LM Test*

Chance-based model risks occur when different model fits arise from different samples of the same population, influenced by factors such as sample size, model complexity, and degrees of freedom. Significant chance-based model risks imply a higher likelihood of overlooking pathways within the model. To illustrate, Figure 4 visualizes the relationship between univariate LM tests, parameters, and various sample sizes. The x-axis represents the distribution of testing parameters, derived from 500 Monte Carlo simulations based on the population and analysis models depicted in Figures 2 and 3. On the other hand, the y-axis illustrates the univariate LM test statistics. Figure 4 highlights the true missing parameters. As observed, when the sample size is small, the distributions of LM tests for all parameters exhibit relatively large variations, including the true missing parameters. This phenomenon arises due to the small sample size relative to the model size and degrees of freedom. Consequently, the standard LM test faces challenges in distinguishing the true missing parameters from other parameters or effectively identifying any potential missing paths, presenting an issue due to its vulnerability to chance-based interpretations. However, as the sample size increases, the variation in the LM tests for the true missing parameters diminishes. This leads to a clearer distinction between the true missing parameters and other parameters, reducing the likelihood of being misled by noise.

Figure 4. Univariate LM test statistics across varying sample sizes

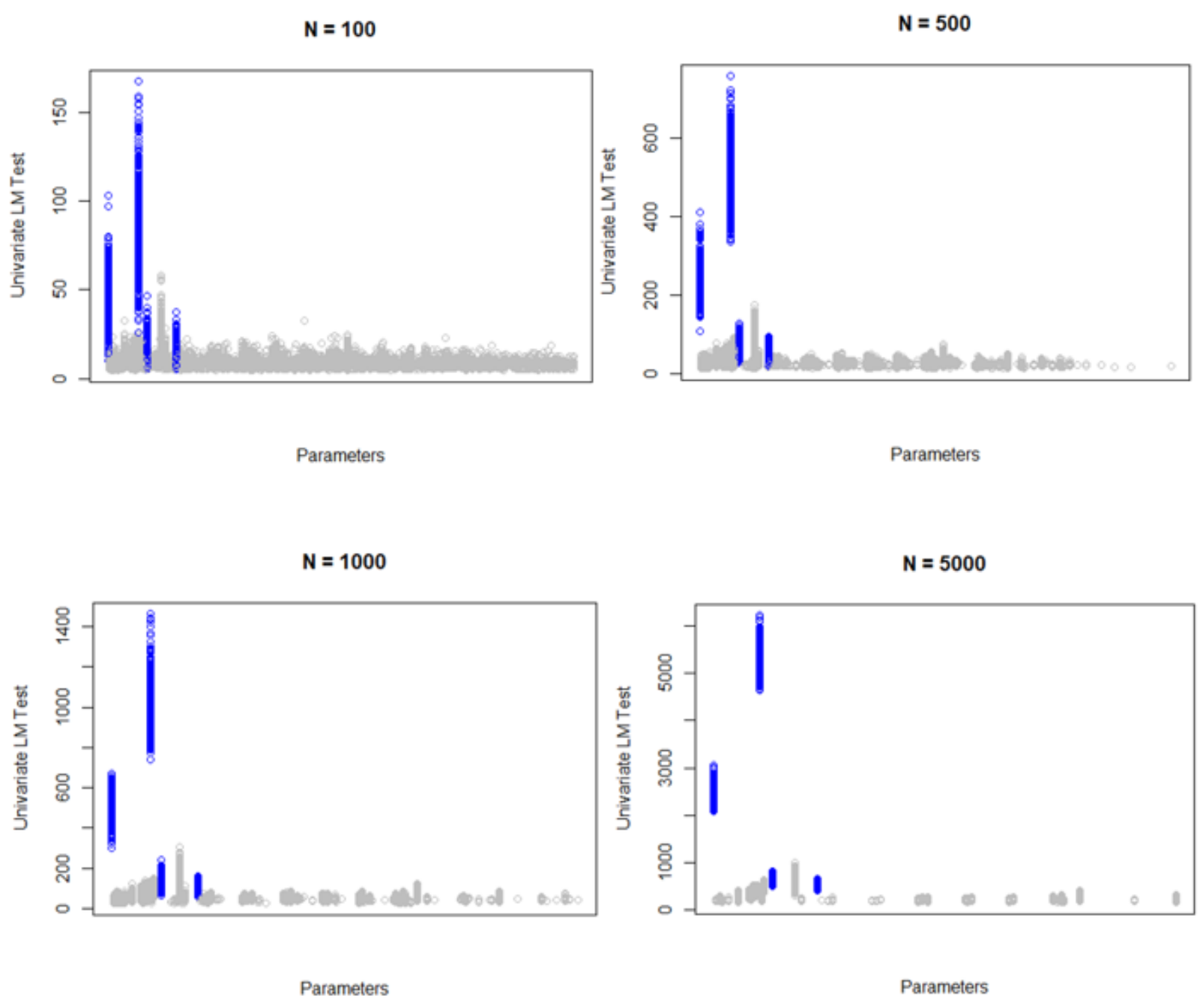

In the third stage of our analysis, we employ a forward stepwise approach along with bootstrap resampling to identify the optimal specifications. While statistically significant large LM test values are observed, they don't necessarily imply that the suggested parameters accurately reflect the 'true' values. This is because the data is generally a sample drawn from the larger population. Furthermore, in real-world data analysis, researchers frequently contend with finite sample sizes and unknown data distributions, giving rise to sample-specific errors and characteristics that may impact the model's accuracy in reflecting the population. Consequently, the parameters recommended by the standard LM test might not generalize effectively to different samples. In this regard, bootstrap resampling can handle the issue of unknown distribution and provide more accurate results than those based on standard asymptotic properties.

In our approach, we use bootstrap resampling in every forward stepwise procedure. The LM test for a set of omitted parameters can be broken down into a series of 1-$df$ tests. Bentler and Dijkstra (1985) developed a forward stepwise LM procedure, where at each step, the parameter is chosen that will maximally increase the LM $\chi^2$. We will perform the bootstrap LM test by examining two parameters at a time. At each step, we randomly draw 500 samples with replacements and compute the mean LM test statistic and its *p*-value. We will repeat this process by adding another pair of parameters until we have tested all possible omitted parameters in the analysis model.

It is crucial to perform multiple repetitions of the test during this process as the LM $\chi^2$ value can vary depending on the model parameters and their correlations with each other. Adding or removing parameters will change the covariance structure for the LM test, and hence, the LM test statistic at each step will provide more accurate information about the remaining missing parameters. We designate the parameters with *p*-values less than 0.05 as statistically significant and refer to them as bootstrap LM-based parameters for the Wald test selection.

### *4.6 Bootstrap Wald Test*

The bootstrap LM tests establish a set of parameters for subsequent validation, while the Wald tests incorporate these recommended parameters and conduct a series of backward stepwise bootstrap Wald tests. The Wald tests assess whether each initially treated-as-free parameter can be collectively set to zero without a significant loss in model fit. This simplifies the model by removing nonsignificant parameters and provides further validation. Parameters that are truly missing exhibit $p$-values < 0.05, confirming their significance and justifying their inclusion in the model. Conversely, parameters that should be excluded from the model yield $p$-values $\geq$ 0.05. This integrated approach effectively addresses the problem of false positives.

## 5 Simulation Results

To evaluate the performance and reliability of the improved LM test, we compare the results across various sample sizes while maintaining consistent degrees of freedom. The likelihood ratio test (LRT) is widely regarded as one of the most commonly used methods for assessing the performance of nested models. In this study, we compare the performances of the improved LM test to LRT for each sample size (see the Appendix for details on LRT calculation). Table 1 shows consistent performance for all models using bootstrap Wald tests across different sample sizes, affirming the combined methodology's effectiveness. The first column displays the top 12 possible parameters based on univariate LM tests, with the highlighted gray parameters representing known omitted parameters. For brevity, the middle two columns only present the bootstrap Wald test (B-Wald) $\chi^2$ statistics and their associated $p$-values. The last two columns show the LRTs and their associated $p$-values. Statistically significant values are highlighted in gray, indicating that their corresponding parameters should be included in the modified model. For detailed test results, please refer to Table A1 in the Appendix. As shown in Table 1, the improved LM tests accurately identify omitted parameters across all sample sizes, consistently outperforming LRTs, especially when sample sizes are small.

Table 1. Test Statistics by Different Sample Sizes

**N=100**

| | Para. | B-Wald | P-value | LRT | P-value |
|---|---|---|---|---|---|
| 1 | **F2, x6** | **63.877** | **0.000** | **363.190** | **0.000** |
| 2 | **F3, 16** | **19.996** | **0.000** | **337.110** | **0.000** |
| 3 | **F2, x20** | **27.681** | **0.000** | **313.420** | **0.000** |
| 4 | **F1, x9** | **19.313** | **0.001** | **283.980** | **0.000** |
| 5 | x4, x8 | | | 283.320 | 0.416 |
| 6 | F2, x2 | 6.206 | 0.094 | **277.230** | **0.014** |
| 7 | x4, x12 | 5.722 | 0.099 | **272.450** | **0.029** |
| 8 | x4, x7 | | | **267.480** | **0.026** |
| 9 | x5, x8 | | | 264.640 | 0.092 |
| 10 | x12, x15 | 5.020 | 0.133 | **259.810** | **0.028** |
| 11 | x6, x4 | 7.359 | 0.081 | **254.870** | **0.026** |
| 12 | F2, x3 | | | 254.620 | 0.620 |

**N=150**

| | Para. | B-Wald | P-value | LRT | P-value |
|---|---|---|---|---|---|
| 1 | **F2, x6** | **115.089** | **0.000** | **375.610** | **0.000** |
| 2 | **F1, x9** | **40.894** | **0.000** | **309.720** | **0.000** |
| 3 | **F3, x16** | **12.019** | **0.006** | **285.180** | **0.000** |
| 4 | **F2, x20** | **39.441** | **0.000** | **241.480** | **0.000** |
| 5 | F3, x9 | 1.056 | 0.474 | 241.480 | 0.931 |
| 6 | F2, x2 | 1.066 | 0.483 | 240.800 | 0.412 |
| 7 | F1, x16 | 4.279 | 0.176 | 237.060 | 0.053 |
| 8 | x6, x4 | 6.939 | 0.106 | **232.230** | **0.028** |
| 9 | x6, x12 | 1.586 | 0.405 | 231.400 | 0.363 |
| 10 | F2, x5 | 1.031 | 0.495 | 230.880 | 0.469 |
| 11 | F2, x8 | 3.972 | 0.230 | 228.680 | 0.138 |
| 12 | F3, x6 | 1.786 | 0.343 | 227.400 | 0.258 |

**N=250**

| | Para. | B-Wald | P-value | LRT | P-value |
|---|---|---|---|---|---|
| 1 | **F2, x6** | **128.070** | **0.000** | **474.880** | **0.000** |
| 2 | **F1, x9** | **71.814** | **0.000** | **387.540** | **0.000** |
| 3 | F3, x6 | 6.920 | 0.066 | 384.990 | 0.111 |
| 4 | **F2, x20** | **63.686** | **0.000** | **312.610** | **0.000** |
| 5 | **F3, x16** | **29.380** | **0.000** | **258.990** | **0.000** |
| 6 | F2, x8 | 1.606 | 0.385 | 258.120 | 0.351 |
| 7 | x6, x20 | 1.386 | 0.434 | 257.880 | 0.622 |
| 8 | F2, x1 | 1.450 | 0.411 | 256.610 | 0.260 |
| 9 | F2, x3 | 0.983 | 0.495 | 256.610 | 0.999 |
| 10 | F1, x13 | 4.793 | 0.149 | 253.160 | 0.063 |
| 11 | x3, x12 | 6.919 | 0.075 | **246.850** | **0.012** |
| 12 | F2, x5 | 0.969 | 0.503 | 246.780 | 0.795 |

**N=350**

| | Para. | B-Wald | P-value | LRT | P-value |
|---|---|---|---|---|---|
| 1 | **F2, x6** | **203.534** | **0.000** | **584.870** | **0.000** |
| 2 | **F1, x9** | **133.946** | **0.000** | **419.970** | **0.000** |
| 3 | **F2, x20** | **88.283** | **0.000** | **291.160** | **0.000** |
| 4 | F3, x6 | 3.770 | 0.176 | **281.600** | **0.002** |
| 5 | x6, x20 | 3.544 | 0.207 | 279.110 | 0.115 |
| 6 | **F3, x16** | **31.355** | **0.000** | **231.680** | **0.000** |
| 7 | F2, x4 | 1.589 | 0.383 | 231.600 | 0.781 |
| 8 | x11, x15 | **11.806** | **0.012** | **219.580** | **0.001** |
| 9 | F2, x3 | 2.146 | 0.306 | 218.730 | 0.358 |
| 10 | F2, x5 | 1.797 | 0.367 | 217.980 | 0.387 |
| 11 | F2, x8 | 1.161 | 0.483 | 217.680 | 0.581 |
| 12 | F2, x2 | 2.280 | 0.287 | 215.920 | 0.185 |

**N=400**

| | Para. | B-Wald | P-value | LRT | P-value |
|---|---|---|---|---|---|
| 1 | **F2, x6** | **214.186** | **0.000** | **537.270** | **0.000** |
| 2 | **F1, x9** | **115.641** | **0.000** | **365.330** | **0.000** |
| 3 | **F2, x20** | **80.792** | **0.000** | **252.840** | **0.000** |
| 4 | **F3, x16** | **80.792** | **0.000** | **208.280** | **0.000** |
| 5 | F2, x5 | 2.363 | 0.290 | 205.690 | 0.108 |
| 6 | F1, x20 | 3.365 | 0.233 | 203.600 | 0.148 |
| 7 | F2, x4 | 2.504 | 0.284 | 201.820 | 0.182 |
| 8 | x20, x11 | 2.073 | 0.360 | 199.950 | 0.171 |
| 9 | x5, x10 | 6.893 | 0.098 | **194.350** | **0.018** |
| 10 | x5, x8 | 2.779 | 0.279 | 192.380 | 0.160 |
| 11 | F1, x11 | 2.092 | 0.338 | 190.940 | 0.229 |
| 12 | F2, x8 | 1.052 | 0.490 | 190.930 | 0.925 |

**N=500**

| | Para. | B-Wald | P-value | LRT | P-value |
|---|---|---|---|---|---|
| 1 | **F2, x6** | **302.512** | **0.000** | **720.280** | **0.000** |
| 2 | **F1, x9** | **189.343** | **0.000** | **466.390** | **0.000** |
| 3 | **F2, x20** | **120.855** | **0.000** | **321.340** | **0.000** |
| 4 | F3, x6 | 1.113 | 0.486 | 321.020 | 0.570 |
| 5 | **F3, x16** | **46.686** | **0.000** | **255.710** | **0.000** |
| 6 | F2, x5 | 2.420 | 0.275 | 254.170 | 0.215 |
| 7 | F3, x9 | 3.392 | 0.238 | 251.400 | 0.096 |
| 8 | F2, x8 | 0.898 | 0.500 | 251.340 | 0.813 |
| 9 | F2, x1 | 1.162 | 0.449 | 251.250 | 0.763 |
| 10 | F2, x3 | 1.476 | 0.414 | 250.680 | 0.451 |
| 11 | x6, x20 | 1.840 | 0.378 | 249.740 | 0.332 |
| 12 | F1, x10 | 5.693 | 0.106 | **244.640** | **0.024** |

**N=700**

| | Para. | B-Wald | P-value | LRT | P-value |
|---|---|---|---|---|---|
| 1 | **F2, x6** | **437.878** | **0.000** | **794.400** | **0.000** |
| 2 | **F1, x9** | **183.251** | **0.000** | **561.740** | **0.000** |
| 3 | **F2, x20** | **187.452** | **0.000** | **321.070** | **0.000** |
| 4 | F3, x6 | 1.177 | 0.484 | 319.520 | 0.214 |
| 5 | F1, x20 | 1.102 | 0.470 | 319.460 | 0.800 |
| 6 | F2, x8 | 3.063 | 0.240 | **315.180** | **0.039** |
| 7 | **F3, x16** | **60.078** | **0.000** | **240.780** | **0.000** |
| 8 | x6, x20 | 2.224 | 0.326 | 239.580 | 0.273 |
| 9 | F2, x4 | 1.181 | 0.454 | 239.500 | 0.787 |
| 10 | F2, x3 | 1.630 | 0.416 | 238.640 | 0.353 |
| 11 | F2, x5 | 1.055 | 0.480 | 238.400 | 0.624 |
| 12 | x6, x8 | 1.829 | 0.401 | 237.680 | 0.397 |

**N=1000**

| | Para. | B-Wald | P-value | LRT | P-value |
|---|---|---|---|---|---|
| 1 | **F2, x6** | **622.506** | **0.000** | **1231.510** | **0.000** |
| 2 | **F1, x9** | **312.393** | **0.000** | **752.360** | **0.000** |
| 3 | **F2, x20** | **265.379** | **0.000** | **407.060** | **0.000** |
| 4 | F3, x6 | 1.257 | 0.455 | 404.750 | 0.129 |
| 5 | **F3, x16** | **97.400** | **0.000** | **266.690** | **0.000** |
| 6 | F2, x8 | 1.369 | 0.460 | 266.590 | 0.761 |
| 7 | F1, x20 | 1.125 | 0.477 | 266.590 | 0.929 |
| 8 | x6, x20 | 3.020 | 0.270 | 264.710 | 0.171 |
| 9 | F2, x5 | 1.188 | 0.471 | 264.710 | 0.940 |
| 10 | F2, x3 | 1.042 | 0.484 | 264.330 | 0.537 |
| 11 | F2, x4 | 1.862 | 0.402 | 263.740 | 0.445 |
| 12 | F2, x1 | 1.403 | 0.447 | 263.600 | 0.702 |

**N=2000**

| | Para. | B-Wald | P-value | LRT | P-value |
|---|---|---|---|---|---|
| 1 | **F2, x6** | **1212.599** | **0.000** | **2087.210** | **0.000** |
| 2 | **F1, x9** | **598.461** | **0.000** | **1240.160** | **0.000** |
| 3 | **F2, x20** | **528.742** | **0.000** | **538.860** | **0.000** |
| 4 | F1, x20 | 1.541 | 0.427 | 538.500 | 0.552 |
| 5 | **F3, x16** | **232.153** | **0.000** | **224.930** | **0.000** |
| 6 | F2, x8 | 1.190 | 0.460 | 224.840 | 0.755 |
| 7 | F3, x6 | 1.196 | 0.479 | 224.750 | 0.766 |
| 8 | x6, x20 | 1.003 | 0.495 | 224.650 | 0.751 |
| 9 | F2, x3 | 1.028 | 0.487 | 224.050 | 0.438 |
| 10 | F2, x5 | 1.995 | 0.374 | 224.020 | 0.887 |
| 11 | F2, x1 | 0.946 | 0.510 | 222.330 | 0.192 |
| 12 | F2, x4 | 9.140 | 0.059 | **215.660** | **0.010** |

## 5.1 *Model Specification Stability and Model Fit Validity*

In this section, we aim to assess the stability of the improved LM test-suggested model fit over repeated samples of different sample sizes. A model that fits the data well should follow a standard $\chi^2$ distribution, $T_{ML} \xrightarrow{\mathcal{L}} \chi^2_{df}$, as $N$ grows larger, demonstrating asymptotic properties (Browne 1984; Jöreskog and Sörbom 1988; Bentler and Dijkstra 1985). Based on this reasoning, we fit the improved LM test-suggested model to the simulated data drawn from the population model (Figure 2). If the $T_{ML}$ are close to the degrees of freedom, it provides strong empirical evidence that the improved LM test-suggested model fits better. To ensure its generalizability, we randomly draw samples of different sizes from the population model and fit the improved LM test-suggested model. If consistency is maintained, we are confident that the improved LM test-suggested model is adequate for general use.

Table 2. Monte Carlo Simulation Results for Asymptotic Properties

| | | | | | NFI | | CFI | | RMSEA | |
|---|---|---|---|---|---|---|---|---|---|---|
| N | $\chi^2$ | SD | P-value | Rej. Rate | 2.5% | 97.5% | 2.5% | 97.5% | 2.5% | 97.5% |
| 100 | 275.650 | 25.824 | 0.184 | 0.386 | 0.748 | 0.836 | 0.922 | 1.000 | 0.000 | 0.059 |
| 150 | 263.874 | 22.901 | 0.277 | 0.205 | 0.829 | 0.885 | 0.960 | 1.000 | 0.000 | 0.042 |
| 200 | 259.340 | 24.503 | 0.332 | 0.167 | 0.866 | 0.913 | 0.970 | 1.000 | 0.000 | 0.036 |
| 250 | 255.329 | 22.978 | 0.371 | 0.115 | 0.894 | 0.929 | 0.978 | 1.000 | 0.000 | 0.031 |
| 300 | 253.984 | 23.378 | 0.394 | 0.109 | 0.911 | 0.939 | 0.982 | 1.000 | 0.000 | 0.029 |
| 400 | 251.213 | 22.787 | 0.424 | 0.095 | 0.933 | 0.954 | 0.988 | 1.000 | 0.000 | 0.023 |
| 500 | 249.332 | 22.811 | 0.442 | 0.078 | 0.946 | 0.963 | 0.991 | 1.000 | 0.000 | 0.020 |
| 800 | 247.594 | 22.136 | 0.467 | 0.063 | 0.966 | 0.976 | 0.994 | 1.000 | 0.000 | 0.015 |
| 1,000 | 247.450 | 22.108 | 0.466 | 0.052 | 0.973 | 0.981 | 0.996 | 1.000 | 0.000 | 0.014 |
| 2,000 | 247.235 | 22.054 | 0.468 | 0.062 | 0.986 | 0.990 | 0.998 | 1.000 | 0.000 | 0.010 |
| 5,000 | 246.822 | 22.195 | 0.477 | 0.062 | 0.994 | 0.996 | 0.999 | 1.000 | 0.000 | 0.006 |
| 10,000 | 245.462 | 22.410 | 0.495 | 0.054 | 0.997 | 0.998 | 1.000 | 1.000 | 0.000 | 0.004 |

Expected mean $\chi^2$ test statistic is 245, expected mean standard deviation is 22.136.

The Monte Carlo simulations in this study are based on Equation 13. We conduct 1,000 trials and calculate the average statistics, which are reported in Table 2. We examine the performance of the analysis model suggested by the improved LM test by varying the sample sizes from 100 to 10,000. Since the simulated data are normally distributed, the maximum likelihood estimator is sufficient to examine the basic statistical performance and asymptotic properties.

Table 2 presents the mean $\chi^2$ test statistics, their mean standard deviations, mean *p*-values, mean rejection rates, and the 2.5th and 97.5th percentiles of fit indices (NFI, CFI, and RMSEA) by sample size. As shown in Table 2, as the sample size increases, all mean statistics test statistics get closer to the expected values: $\chi^2$=245, SD=22.136, *p*-value=0.50, and empirical rejection rate is 0.05. Note that when sample sizes are less than 500, the $\chi^2$ test statistics are increasingly inflated, deviating from the expected value of 245. As documented by previous studies, ML estimator is biased against small sample sizes (Hayakawa 2019; Arruda and Bentler 2017; Zheng and Bentler 2021, 2023).

In addition, NFI and CFI become closer to 1, and RMSEA is about 0 when the sample sizes are greater than 200. The collective statistical indicators provide compelling evidence that the improved LM test effectively detected the omitted parameters in the analysis model and provided a satisfactory fit to the simulated data. Moreover, the modified model, derived from the specification search results using the improved LM test, exhibited a robust statistical fit across various sample sizes.

### *5.2 Robustness of the Improved LM Test*

To test the robustness of the improved LM test, we vary the magnitudes of factor correlations, the number of indicators per factor from low to high, and factor loadings. First, we find that with more indicators per factor, it becomes easier to detect omitted parameters. When the number of indicators per factor is fewer, the statistical power of the improved LM test is weakened, particularly with smaller sample sizes. Nevertheless, the improved LM test still outperforms the LRT across all sample sizes. When the number of indicators per factor increases, both the improved LM test and LRT perform similarly.

Second, when factor correlations are low, the improved LM test becomes more efficient at detecting omitted parameters. The performances of the improved LM test and the LRT become similar when sample sizes exceed 100. However, with smaller sample sizes, the improved LM test consistently outperforms the LRT. In contrast, when factor correlations are high, increasing potential relationships among factor loadings and residuals, the improved LM test continues to deliver outstanding performance.

Third, the magnitudes of factor loadings influence the performance of the improved LM test, and this is dependent on the sample size. When $N \geq 400$, the improved LM test delivers efficient and robust performance compared to the LRT. However, low factor loadings in smaller sample sizes tend to have a stronger impact on the detection of omitted variables and convergence. We found that when N < 400, the models encounter convergence issues, mainly because the covariance matrix becomes not positive definite. In contrast, with high factor loadings, both the improved LM test and LRT perform well across all sample sizes in this study. Nonetheless, the improved LM test consistently outperforms the LRT in detecting correct parameters. For the results of the simulation tests, please refer to the Appendix.

# 6 Empirical Examples

## *6.1 Example 1. National Identity and Patriotism*

In this study, we evaluate the effectiveness of the improved LM test using a covariance structure model constructed from Huddy and Khatib's (2007) student data, gathered in 2002. For detailed data collection and student sample information, please refer to page 66 in Huddy & Khatib (2007). This dataset comprises 341 respondents. The survey questions use a 4-point Likert scale, with response options ranging from 'strongly approve' to 'strongly disapprove.' We employ the *diagonally weighted least squares* (DWLS) estimator to handle the ordered categorical variables. For brevity, the indicators and factors are unlabeled here; please refer to the appendix for the survey questions.

Figure 5. Path diagram of National Identity and Patriotism (Huddy and Khatib 2007)

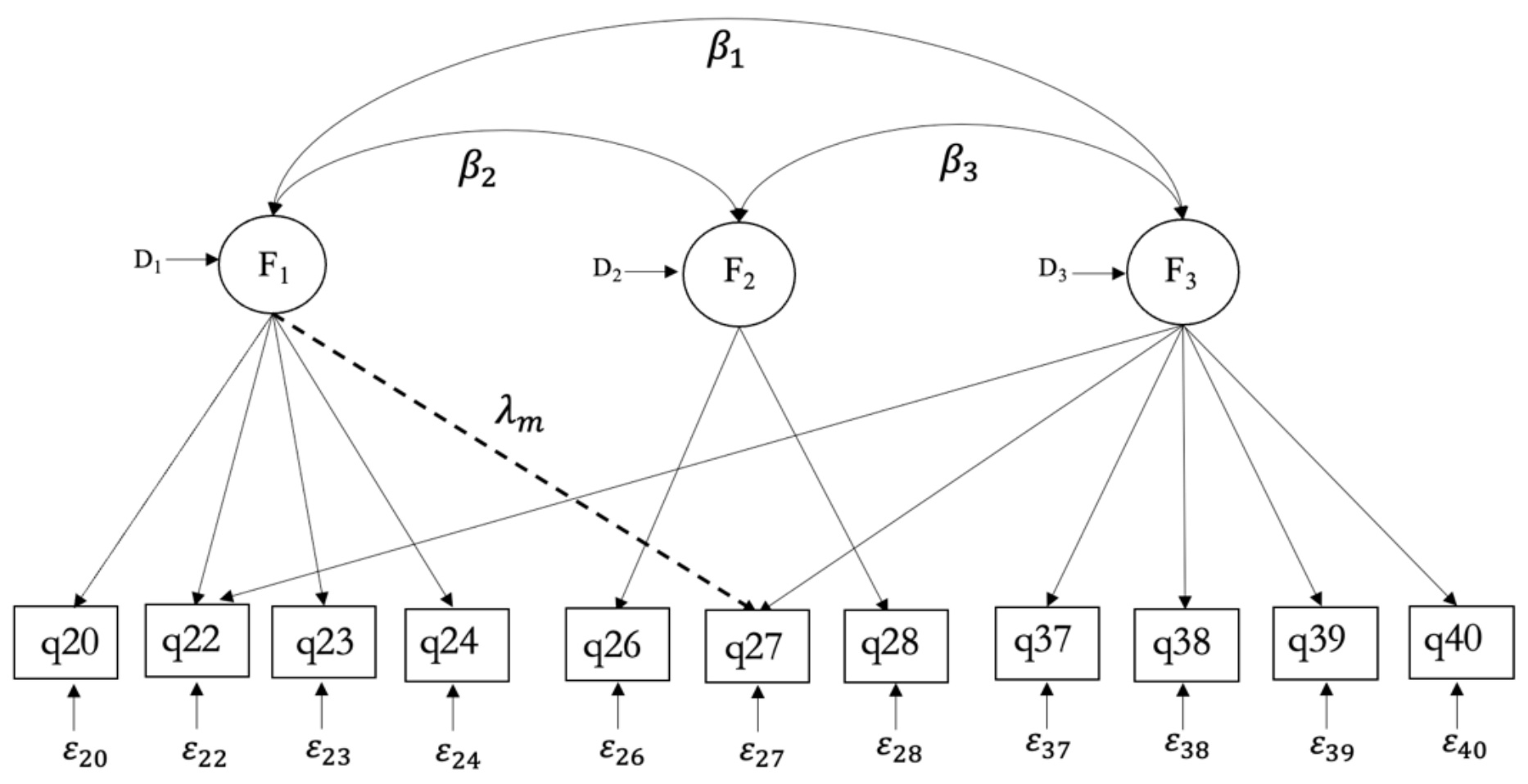

The path diagram for the three-factor model is depicted in Figure 5. This model involves a small sample size (N=341) relative to a larger number of degrees of freedom ($df$=39). The national identity and patriotism model consists of three latent factors, $F_1$, $F_2$, and $F_3$. These factors represent the constructs of *national identity*, *symbolic patriotism*, and *uncritical patriotism*, respectively. Each of these constructs is assessed by a series of indicators $X_i$. Factor loadings are denoted by $\lambda_i$, residuals by $\varepsilon_i$, and the residuals of factor by $D_i$. The coefficients $\beta_1$, $\beta_2$ and $\beta_3$ measure the correlations between the three factors. To evaluate the performance of the improved LM test, we follow the same procedure as employed in the simulated data. The dashed lines in Figure 5 represent the recommended parameters suggested by the improved LM test.

Table 3. Summary of Example 1 Test Statistics

| | | LM test | | Bootstrap LM Test | | Bootstrap W Test | |
|---|---|---|---|---|---|---|---|
| | Parameters | LM $\chi^2$ | P-values | LM $\chi^2$ | P-values | Chi-square | P-values |
| 1 | F2, q27 | 22.291 | 0.000 | **14.221** | **0.004** | 5.948 | 0.081 |
| 2 | **F1, q27** | 19.856 | 0.000 | **13.102** | **0.005** | **7.628** | **0.041** |
| 3 | q27, q38 | 10.340 | 0.001 | **6.761** | **0.023** | 1.554 | 0.392 |
| 4 | q37, q38 | 10.311 | 0.001 | **6.670** | **0.035** | 0.497 | 0.601 |
| 5 | F1, q40 | 9.749 | 0.002 | **6.829** | **0.038** | 2.206 | 0.225 |
| 6 | q40, q38 | 7.495 | 0.006 | 4.982 | 0.068 | | |
| 7 | q40, q39 | 6.701 | 0.010 | 5.166 | 0.106 | | |
| 8 | F2, q37 | 6.247 | 0.012 | 4.240 | 0.077 | | |
| 9 | q37, q39 | 6.231 | 0.013 | 4.423 | 0.109 | | |
| 10 | q27, q39 | 5.663 | 0.017 | 3.808 | 0.088 | | |

Table 3 displays the outcomes of three distinct tests: the univariate LM test, bootstrap LM test, and bootstrap Wald test. The bootstrap LM test, executed through a forward stepwise approach, identified five missing parameters that were statistically significant. The highlighted items indicate the actual missing parameters, and the items in bold in the bootstrap LM and Wald tests are statistically significant. The bootstrap Wald test concurred that the parameter $\lambda_m$ (the factor loading linking $F_1$ and q27) should be included in the original model to improve the model fit. q27 inquires, "How angry does it make you feel, if at all, when you hear someone criticizing the United States?" Response options range from extremely angry to not at all. Furthermore, the standardized factor loading of $\lambda_m$ is 0.47 and statistically significant. This indicates that differing levels of national identity ($F_1$) and uncritical patriotism ($F_3$) are likely to influence feelings of anger. While the addition of this parameter may not alter the overall substantive conclusion, it provides new insights into the nuances of these latent structural relationships. However, without including this parameter, Huddy and Khatib's (2007) original model suffers from some degree of misspecification.

Table 4. Comparison of Test Statistics and Model Fit in Example 1

| | Original | Improved LM | Differences |
|---|---|---|---|
| Chi-square | 65.176 | 37.642 | 27.534 |
| Degrees of freedom | 40 | 39 | 1 |
| $P$-value | 0.007 | 0.532 | -0.525 |
| NFI | 0.992 | 0.996 | -0.004 |
| CFI | 0.997 | 1.000 | -0.003 |
| TLI | 0.996 | 1.000 | -0.004 |
| RMSEA | 0.043 | 0.000 | 0.043 |

Note: The statistics for the original model are obtained from a replication of the initial model. While these statistics may not be identical to the original test results, they are close to them.

Table 4 presents the results of our replication study based on Huddy & Khatib's (2007) research using the DWLS estimator. In the original study, the $\chi^2$ statistic is 65.176 with 40 degrees of freedom, resulting in a *p*-value of 0.007. The CFI is 0.997, NFI is 0.992, TLI is 0.996, and the RMSEA is 0.043. The $\chi^2$ test statistic provides limited support for the substantive argument. However, upon introducing the omitted parameter, $\lambda_m$, as suggested by the improved LM test, the $\chi^2$ test statistic reduces to 27.534, resulting in a *p*-value of 0.532. The improvement in the $\chi^2$ test statistic is crucial as it suggests that the model-implied covariance structure is highly consistent with the sample covariance structure. Moreover, the NFI increases to 0.996, the CFI and TLI increase to 1.0, and the RMSEA decreases to 0. The final column in Table 4 indicates the differences in test statistics and fit indices between the two models, showcasing a significant enhancement in model fit and strengthening the theoretical argument based on this structural relationship.

## 6.2 *Example 2: SEM of Relationship of Human Value Priorities*

In another empirical application, we conducted an analysis of the German sample (N=2,919) from the 2002 European Social Survey, which is a cross-national probability survey. To illustrate the effectiveness of our improved LM test in small sample sizes, we randomly selected 500 observations from the German sample. Detailed information on data collection procedures and original survey questions can be found on the ESS website.

To analyze the data, we employed a four-factor SEM model, as depicted in Figure 6, following standard practice. The model comprises four latent factors represented by ovals, each measured by multiple indicators. In the original model, factor $F_3$ is predicted by factors $F_1$ and $F_2$, while factor $F_4$ is predicted by factors $F_1$ and $F_2$. Furthermore, there are correlations between factors $F_1$ and $F_2$, as well as $F_3$ and $F_4$. To evaluate the effectiveness of our proposed improved LM test approach, we removed the coefficient parameters between $F_2$ and $F_3$, and between $F_3$ and $F_4$, as indicated by the dash lines in Figure 6.

Figure 6. SEM of Human Value Priorities (Davidov 2009; Oberski 2014)

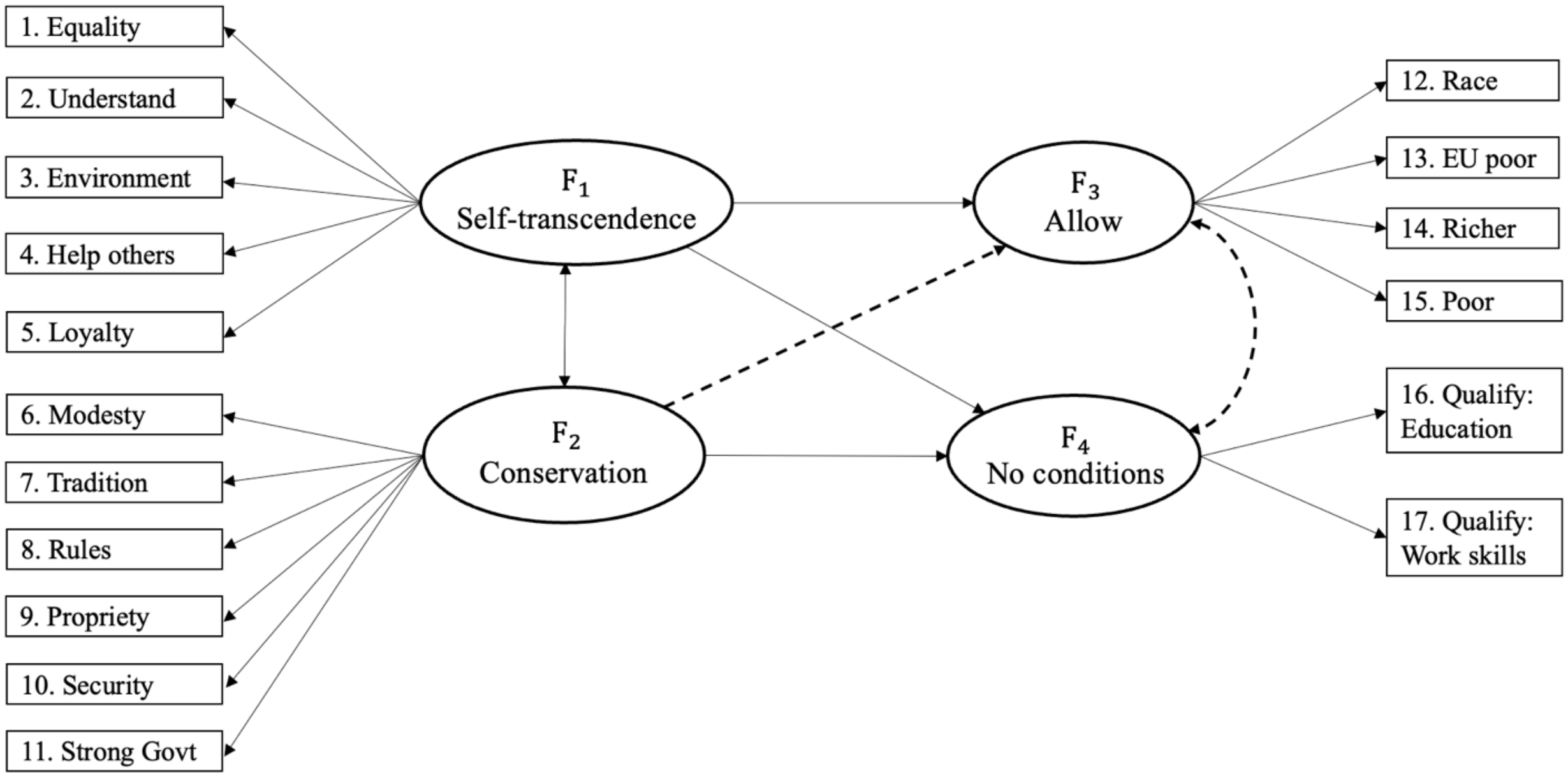

Note: Error and factor variances are not shown in the path diagram.

Table 5. Summary of Example 2 Test Statistics

| | | LM Test | | Bootstrap LM Test | | Bootstrap Wald Test | |
|---|---|---|---|---|---|---|---|
| | Parameters | LM $\chi^2$ | P-values | LM $\chi^2$ | P-values | Chi-square | P-values |
| 1 | **F1, F3** | 64.563 | 0.000 | **63.523** | **0.000** | **32.676** | **0.000** |
| 2 | **F2, F3** | 64.563 | 0.000 | **63.523** | **0.000** | **17.324** | **0.001** |
| 3 | Rules, Propriety | 25.415 | 0.000 | **25.920** | **0.002** | 9.093 | 0.104 |
| 4 | **F3, F4** | 23.890 | 0.000 | **23.690** | **0.000** | **22.676** | **0.001** |
| 5 | **F3, Understand** | 19.611 | 0.000 | **20.327** | **0.003** | **10.543** | **0.043** |
| 6 | **Equality, Tradition** | 14.447 | 0.000 | **15.342** | **0.007** | **8.826** | **0.049** |
| 7 | F2, Richer | 13.483 | 0.000 | **13.698** | **0.012** | 1.462 | 0.423 |
| 8 | F1, Rules | 12.185 | 0.000 | **13.063** | **0.015** | 0.931 | 0.512 |
| 9 | Environment, Tradition | 11.458 | 0.001 | **12.283** | **0.026** | 4.348 | 0.195 |
| 10 | F2, Help others | 11.245 | 0.001 | **13.235** | **0.015** | 6.228 | 0.103 |

Table 5 presents the results of three tests conducted on the model: the LM test, bootstrap LM test, and bootstrap Wald test. The LM test identified the top 10 omitted parameters based on LM test statistics, out of which two parameters ($F_2$, $F_3$) and ($F_3$, $F_4$) were the actual missing parameters. The improved LM tests validated these two parameters and suggested two more parameters ($F_3$, Understand) and (Equality, Tradition) based on the ten testing parameters in the model. Note that ($F_1$, $F_3$) is meaningless here because the original research aims to use a unidirectional arrow to indicate the effect of $F_1$ on $F_3$ based on their theoretical argument, while the improved LM test suggests a correlation instead. Thus, we disregard this suggested omitted parameter. The suggested parameter (F3, Understand) suggests that attitudes toward immigration may also be influenced by the universalism value, which emphasizes understanding and concern for the welfare of all people, as well as the influence of self-transcendence. Additionally, the suggested parameter (Equality, Tradition) indicates that the belief in treating every person equally is correlated with the value placed on tradition. Naturally, this relationship may vary significantly across different countries and cultures. Such variations could introduce measurement invariance and model misspecification issues when comparing the effects of values on attitudes toward immigration without accounting for these structural relationships.

To determine whether adding these suggested parameters could improve the model fit, we compared the test statistics and fit indices. Table 6 reports that the original model's $\chi^2$ test is 324.224 with 115 degrees of freedom. However, when we added all the suggested parameters to the model, the $\chi^2$ test decreased to 236.204, a reduction of 87.822, with a loss of only 3 degrees of freedom. Additionally, the NFI increased from 0.844 to 0.886, the CFI increased from 0.892 to 0.936, and the RMSEA decreased from 0.064 to 0.050. All these statistics confirmed that the improved LM test method yielded favorable results for this model by incorporating the additional parameters it recommended.

Table 6. Comparisons of Test Statistics and Fit Indices

| | Original Model | Modified Model | Difference |
|---|---|---|---|
| Chi-square | 324.224 | 236.402 | 87.822 |
| DF | 115 | 112 | 3 |
| P-value | 0.000 | 0.000 | 0 |
| NFI | 0.844 | 0.886 | -0.042 |
| CFI | 0.892 | 0.936 | -0.044 |
| RMSEA | 0.064 | 0.050 | 0.014 |

In summary, the improved LM test effectively identified omitted parameters in empirical examples 1 and 2. Incorporating these parameters, as shown in Tables 4 and 6, substantially improves the overall model fit in the $\chi^2$ test statistics and fit indices. While these additions may not alter the substantive conclusions, they enhance confidence in the authors' arguments and provide new insights into their theoretical claims. It is important to note, however, that although the improved LM test is a valuable data-driven method for uncovering hidden parameters, the decision to include suggested parameters should be guided by strong theoretical justification.

## 7 Conclusion

CSA and SEM stand as formidable tools that enjoy wide adoption in the behavioral and social sciences, facilitating the understanding of latent structural relationships among variables. Nevertheless, the creation of an accurate model proves challenging, and conventional practices occasionally engender the perils of chance-based model risks occur when different model fits arise from different samples of the same population, influenced by factors such as sample size, model complexity, and degrees of freedom. Significant chance-based model risks imply a higher likelihood of overlooking pathways within the model and undue rejection of the null hypothesis. To surmount these predicaments, the present study proposed an improved LM test, designed to rectify instances of falsely statistically significant parameters and to effectively pinpoint omitted parameters, particularly in scenarios featuring modest sample sizes. The improved LM test integrates bootstrap LM and Wald tests, enhancing model specification searches by accurately identifying missing parameters. This robust framework advances the field, enabling researchers to effectively model complex phenomena and make well-informed decisions based on well-specified models.

Though our investigation predominantly centers on a model boasting a substantial number of degrees of freedom, it is reasonable to anticipate that our approach will likewise prove efficacious for models featuring fewer degrees of freedom. This expectation stems from the recognition that a model with fewer degrees of freedom reduces the likelihood of succumbing to the perils of chance-based model risks, which occur when different model fits arise from different samples of the same population, influenced by factors such as sample size, model complexity, and degrees of freedom. Significant chance-based model risks imply a higher likelihood of overlooking pathways within the model. Our confidence in the applicability of this approach is reinforced by replicating empirical examples, such as Huddy and Khatib's (2007) model of national identity and patriotism, which involved a small sample size relative to a larger number of degrees of freedom. Similarly, Davidov's (2009) and Oberski's (2014) SEM models examining the relationship of human value priorities exhibit a moderate sample size accompanied by a comparatively greater number of degrees of freedom.

Our simulations, which varied the magnitudes of factor loadings, factor correlations, and the number of indicators per factor, consistently showed that our proposed improved LM test performs noticeably better than the LRT. These extensive evaluations under various realistic scenarios provided further insight into the application and effectiveness of the improved LM test. Nevertheless, the simulation study conducted should not be considered as a comprehensive evaluation encompassing a broad spectrum of realistic conditions. Similarly, conducting a systematic comparison of the practical utility of the improved LM test with other testing methods across diverse topics, such as nonnormal data, varying levels of model complexity, degrees of misspecification, and so on, was beyond the scope of this specific study. However, we remain confident in the contributions made by this research. Future studies could further explore the potential applications and comparative effectiveness of the improved LM test. Additionally, this test is not limited to CSA and SEM; future research could expand its use to regression and other domains, offering broader applicability for applied researchers.

Lastly, as numerous scholars have rightly emphasized, researchers bear the crucial responsibility of interpreting the results yielded by any proposed approach with caution, ensuring they remain aligned with substantive theory (MacCallum, Roznowski, and Nectowitz 1992; Bentler 2006). The improved LM test is no exception. It is vital to recognize that the results of the improved LM test should be considered merely as a suggestion for including statistically indispensable parameters. The decision to integrate these parameters into a modified model should be guided by a solid theoretical foundation.

**Statements and Declarations**

This manuscript has not been submitted for review elsewhere. The research was conducted in adherence to the Ethical Principles of *Political Science Research and Methods* and Code of Conduct, following all appropriate ethical guidelines. No funding was received for the implementation of this study. The authors have no relevant financial or non-financial interests to declare.

# Online Appendix

## Table of Contents

## Likelihood Ratio Test

The likelihood ratio test (LRT) is widely regarded as one of the most commonly used methods for assessing the performance of nested models. In this study, we compare our proposed improved LM test to LRT. LRT can be done as follows. One would create a restricted model $M_1$ in which all of the free parameters $\boldsymbol{\theta_1}$ to be tested are simultaneously set to zero, assuming these parameters do not contribute to explaining the variance in the data. The more general model $M_2$ has added parameters $\boldsymbol{\theta_r}$, so all the parameters in the general model are $\boldsymbol{\theta_2} = \boldsymbol{\theta_1} + \boldsymbol{\theta_r}$, thus $\boldsymbol{\theta_1}$ is a subset of $\boldsymbol{\theta_2}$.

LRT compares the $\chi^2$ test statistics of two nested models by evaluating the difference in $\chi^2$ values obtained from the two runs, as well as by calculating the corresponding difference in degrees of freedom between the restricted model $M_1$ and the more parameterized model $M_2$. It assesses whether the $\chi^2$ difference, also following a $\chi^2$distribution under the null hypothesis, is statistically significant. This tests the hypothesis that the covariance matrices of the two models are equivalent. If this null hypothesis is rejected, it indicates that the additional parameters $\boldsymbol{\theta_r}$ in the more complex model $M_2$ significantly improve the model's fit to the data. With $q_1$ free parameters in $\boldsymbol{\theta_r}$, $\boldsymbol{\theta_2}$ thus has $q_2$ free parameters, where $q_2 = q_1 - r$. When fitting the same covariance matrix with *p* variables, the degrees of freedom associated with $M_1$ and $M_2$ are $df_1$=($p^* - q_1$) and $df_2$=($p^* - q_2$)=( $p^* - q_1 - r$), respectively. Let $\widehat{\boldsymbol{\theta}}_{\boldsymbol{1}}$ and $\widehat{\boldsymbol{\theta}}_{\boldsymbol{2}}$ represent the estimations of $\boldsymbol{\theta_1}$ and $\boldsymbol{\theta_2}$. The LRT compares the $\chi^2$ test statistics of the two models, denoted as $nF(\hat{\boldsymbol{\theta}}_{\mathbf{1}})$ and $nF(\hat{\boldsymbol{\theta}}_{\mathbf{2}})$ respectively. The degrees of freedom associated with the LRT statistic are $df_1 - df_2$. To simplify, $\boldsymbol{\theta_r}$ is assumed to be fixed at 0 in the $M_1$ model. Hence, to compare the difference between $M_1$ and $M_2$, we test if $\boldsymbol{\theta_r}$ differs from 0. Therefore, the LRT statistic can be formulated as:

$$LRT = n\left[F(\hat{\boldsymbol{\theta}}_{\mathbf{1}}) - F(\hat{\boldsymbol{\theta}}_{\mathbf{2}})\right] \sim \chi_r^2.$$

## How do we fit likelihood ratio test?

First, we based on the univariate LM test to draw 12 suggested omitted parameters, so that we have the same set of testing parameters as the improved LM test. Based on these 12 suggested omitted parameters, we add one of them at a time, that way we create a set of nested models. The likelihood ratio tests compare a pair of models at a time sequentially. If a model is better than the other one, then we will see that $\chi^2$ statistic is statistically significant.

Table A1. Test Statistics by Different Sample Sizes (Full results)

N=100

| | | Univariate LM Test | | Bootstrap LM Test | | Bootstrap W Test | | LRT | |
|---|---|---|---|---|---|---|---|---|---|
| | Parameters | LM test | P-values | LM test | P-values | Chi-square | P-values | Chi-square | P-values |
| 1 | **F2, x6** | 92.098 | 0.000 | 67.646 | 0.000 | **63.877** | **0.000** | **363.19** | **0.000** |
| 2 | **F3, x16** | 24.355 | 0.000 | 23.663 | 0.000 | **19.996** | **0.000** | **337.11** | **0.000** |
| 3 | **F2, x20** | 20.773 | 0.000 | 22.001 | 0.005 | **27.681** | **0.000** | **313.42** | **0.000** |
| 4 | **F1, x9** | 17.747 | 0.000 | 18.401 | 0.007 | **19.313** | **0.001** | **283.98** | **0.000** |
| 5 | x4, x8 | 12.349 | 0.000 | 10.732 | 0.058 | | | 283.32 | 0.416 |
| 6 | F2, x2 | 11.601 | 0.001 | 10.763 | 0.029 | 6.206 | 0.094 | **277.23** | **0.014** |
| 7 | x4, x12 | 11.393 | 0.001 | 10.854 | 0.022 | 5.722 | 0.099 | **272.45** | **0.029** |
| 8 | x4, x7 | 10.763 | 0.001 | 10.934 | 0.021 | | | **267.48** | **0.026** |
| 9 | x5, x8 | 10.058 | 0.002 | 9.124 | 0.057 | | | 264.64 | 0.092 |
| 10 | x12, x15 | 9.999 | 0.002 | 10.155 | 0.034 | 5.020 | 0.133 | **259.81** | **0.028** |
| 11 | x6, x4 | 9.644 | 0.002 | 8.254 | 0.035 | 7.359 | 0.081 | **254.87** | **0.026** |
| 12 | F2, x3 | 8.573 | 0.003 | 7.374 | 0.063 | | | 254.62 | 0.620 |

N=150

| | | Univariate LM Test | | Bootstrap LM Test | | Bootstrap W Test | | LRT | |
|---|---|---|---|---|---|---|---|---|---|
| | Parameters | LM test | P-values | LM test | P-values | Chi-square | P-values | Chi-square | P-values |
| 1 | **F2, x6** | 168.551 | 0.000 | 137.789 | 0.000 | **115.089** | **0.000** | **375.610** | **0.000** |
| 2 | **F1, x9** | 74.895 | 0.000 | 69.773 | 0.000 | **40.894** | **0.000** | **309.720** | **0.000** |
| 3 | **F3, x16** | 30.653 | 0.000 | 28.847 | 0.002 | **12.019** | **0.006** | **285.180** | **0.000** |
| 4 | **F2, x20** | 22.704 | 0.000 | 24.366 | 0.001 | **39.441** | **0.000** | **241.480** | **0.000** |
| 5 | F3, x9 | 22.238 | 0.000 | 19.485 | 0.015 | 1.056 | 0.474 | 241.480 | 0.931 |
| 6 | F2, x2 | 20.780 | 0.000 | 17.325 | 0.003 | 1.066 | 0.483 | 240.800 | 0.412 |
| 7 | F1, x16 | 19.006 | 0.000 | 18.947 | 0.005 | 4.279 | 0.176 | 237.060 | 0.053 |
| 8 | x6, x4 | 16.978 | 0.000 | 13.988 | 0.015 | 6.939 | 0.106 | **232.230** | **0.028** |
| 9 | x6, x12 | 16.882 | 0.000 | 15.180 | 0.009 | 1.586 | 0.405 | 231.400 | 0.363 |
| 10 | F2, x5 | 16.355 | 0.000 | 13.907 | 0.012 | 1.031 | 0.495 | 230.880 | 0.469 |
| 11 | F2, x8 | 16.268 | 0.000 | 12.154 | 0.031 | 3.972 | 0.230 | 228.680 | 0.138 |
| 12 | f3, x6 | 15.809 | 0.000 | 19.202 | 0.040 | 1.786 | 0.343 | 227.400 | 0.258 |

N=250

| | | Univariate LM Test | | Bootstrap LM Test | | Bootstrap W Test | | LRT | |
|---|---|---|---|---|---|---|---|---|---|
| | Parameters | LM test | P-values | LM test | P-values | Chi-square | P-values | Chi-square | P-values |
| 1 | **F2, x6** | 243.499 | 0.000 | 227.021 | 0.000 | **128.070** | **0.000** | **474.88** | **0.000** |
| 2 | **F1, x9** | 101.591 | 0.000 | 92.615 | 0.000 | **71.814** | **0.000** | **387.54** | **0.000** |
| 3 | F3, x6 | 92.413 | 0.000 | 86.370 | 0.000 | 6.920 | 0.066 | 384.99 | 0.111 |
| 4 | **F2, x20** | 57.578 | 0.000 | 57.658 | 0.000 | **63.686** | **0.000** | **312.61** | **0.000** |
| 5 | **F3, x16** | 50.632 | 0.000 | 51.008 | 0.000 | **29.380** | **0.000** | **258.99** | **0.000** |
| 6 | F2, x8 | 25.454 | 0.000 | 23.029 | 0.001 | 1.606 | 0.385 | 258.12 | 0.351 |
| 7 | x6, x20 | 24.994 | 0.000 | 24.250 | 0.001 | 1.386 | 0.434 | 257.88 | 0.622 |
| 8 | F2, x1 | 21.83 | 0.000 | 20.145 | 0.001 | 1.450 | 0.411 | 256.61 | 0.260 |
| 9 | F2, x3 | 21.321 | 0.000 | 19.433 | 0.001 | 0.983 | 0.495 | 256.61 | 0.999 |
| 10 | F1, x13 | 19.161 | 0.000 | 18.067 | 0.002 | 4.793 | 0.149 | 253.16 | 0.063 |
| 11 | x3, x12 | 18.871 | 0.000 | 19.316 | 0.002 | 6.919 | 0.075 | **246.85** | **0.012** |
| 12 | F2, x5 | 18.293 | 0.000 | 16.324 | 0.005 | 0.969 | 0.503 | 246.78 | 0.795 |

N=300

| | | Univariate LM Test | | Bootstrap LM Test | | Bootstrap W Test | | LRT | |
|---|---|---|---|---|---|---|---|---|---|
| | Parameters | LM test | P-values | LM test | P-values | Chi-square | P-values | Chi-square | P-values |
| 1 | **F2, x6** | 267.745 | 0.000 | 246.558 | 0.000 | **182.242** | **0.000** | **514.030** | **0.000** |
| 2 | **F1, x9** | 155.903 | 0.000 | 152.754 | 0.000 | **103.410** | **0.000** | **367.540** | **0.000** |
| 3 | **F2, x20** | 68.739 | 0.000 | 70.415 | 0.000 | **70.999** | **0.000** | **266.730** | **0.000** |
| 4 | F1, x20 | 32.597 | 0.000 | 30.873 | 0.007 | 1.319 | 0.457 | 266.610 | 0.736 |
| 5 | **F3, x16** | 30.947 | 0.000 | 30.117 | 0.000 | **29.494** | **0.000** | **227.190** | **0.000** |
| 6 | f2, x5 | 27.228 | 0.000 | 25.242 | 0.000 | 0.753 | 0.547 | 226.690 | 0.480 |
| 7 | x6, x20 | 26.383 | 0.000 | 28.026 | 0.001 | 1.563 | 0.412 | 226.410 | 0.593 |
| 8 | F2, x8 | 24.648 | 0.000 | 23.483 | 0.001 | 1.039 | 0.477 | 226.410 | 0.986 |
| 9 | F2, x1 | 21.296 | 0.000 | 20.915 | 0.002 | 1.941 | 0.337 | 225.410 | 0.317 |
| 10 | F2, x4 | 21.083 | 0.000 | 20.786 | 0.003 | 1.253 | 0.449 | 225.250 | 0.692 |
| 11 | x8, x3 | 20.953 | 0.000 | 20.128 | 0.004 | 3.119 | 0.267 | 223.340 | 0.167 |
| 12 | x6, x12 | 20.022 | 0.000 | 20.205 | 0.006 | 1.624 | 0.430 | 223.240 | 0.756 |

N=400

| | Parameters | Univariate LM Test | | Bootstrap LM Test | | Bootstrap W Test | | LRT | |
|---|---|---|---|---|---|---|---|---|---|
| | | LM test | P-values | LM test | P-values | Chi-square | P-values | Chi-square | P-values |
| 1 | **F2, x6** | 288.109 | 0.000 | 282.443 | 0.000 | **214.186** | **0.000** | **537.27** | **0.000** |
| 2 | **F1, x9** | 206.289 | 0.000 | 207.344 | 0.000 | **115.641** | **0.000** | **365.33** | **0.000** |
| 3 | **F2, x20** | 73.277 | 0.000 | 73.805 | 0.000 | **80.792** | **0.000** | **252.84** | **0.000** |
| 4 | **F3, x16** | 41.481 | 0.000 | 41.395 | 0.000 | **80.792** | **0.000** | **208.28** | **0.000** |
| 5 | F2, x5 | 28.641 | 0.000 | 29.372 | 0.000 | 2.363 | 0.290 | 205.69 | 0.108 |
| 6 | F1, x20 | 26.982 | 0.000 | 27.045 | 0.005 | 3.365 | 0.233 | 203.6 | 0.148 |
| 7 | F2, x4 | 25.02 | 0.000 | 27.045 | 0.005 | 2.504 | 0.284 | 201.82 | 0.182 |
| 8 | x20, x11 | 24.793 | 0.000 | 25.034 | 0.000 | 2.073 | 0.360 | 199.95 | 0.171 |
| 9 | x5, x10 | 23.098 | 0.000 | 23.713 | 0.001 | 6.893 | 0.098 | **194.35** | **0.018** |
| 10 | x5, x8 | 21.474 | 0.000 | 21.412 | 0.001 | 2.779 | 0.279 | 192.38 | 0.160 |
| 11 | F1, x11 | 21.401 | 0.000 | 22.131 | 0.002 | 2.092 | 0.338 | 190.94 | 0.229 |
| 12 | F2, x8 | 21.241 | 0.000 | 21.832 | 0.002 | 1.052 | 0.490 | 190.93 | 0.925 |

N=500

| | Parameters | Univariate LM Test | | Bootstrap LM Test | | Bootstrap W Test | | LRT | |
|---|---|---|---|---|---|---|---|---|---|
| | | LM test | P-values | LM test | P-values | Chi-square | P-values | Chi-square | P-values |
| 1 | **F2, x6** | 538.391 | 0.000 | 483.535 | 0.000 | **302.512** | **0.000** | **720.28** | **0.000** |
| 2 | **F1, x9** | 328.768 | 0.000 | 306.362 | 0.000 | **189.343** | **0.000** | **466.39** | **0.000** |
| 3 | **F2, x20** | 92.473 | 0.000 | 97.219 | 0.000 | **120.855** | **0.000** | **321.34** | **0.000** |
| 4 | F3, x6 | 86.143 | 0.000 | 90.821 | 0.000 | 1.113 | 0.486 | 321.02 | 0.570 |
| 5 | **F3, x16** | 69.436 | 0.000 | 67.891 | 0.000 | **46.686** | **0.000** | **255.71** | **0.000** |
| 6 | F2, x5 | 53.813 | 0.000 | 46.616 | 0.000 | 2.420 | 0.275 | 254.17 | 0.215 |
| 7 | F3, x9 | 51.487 | 0.000 | 45.364 | 0.010 | 3.392 | 0.238 | 251.4 | 0.096 |
| 8 | F2, x8 | 48.234 | 0.000 | 40.207 | 0.000 | 0.898 | 0.500 | 251.34 | 0.813 |
| 9 | F2, x1 | 45.938 | 0.000 | 40.969 | 0.000 | 1.162 | 0.449 | 251.25 | 0.763 |
| 10 | F2, x3 | 39.532 | 0.000 | 33.452 | 0.000 | 1.476 | 0.414 | 250.68 | 0.451 |
| 11 | x6, x20 | 38.695 | 0.000 | 39.874 | 0.000 | 1.840 | 0.378 | 249.74 | 0.332 |
| 12 | F1, x10 | 35.594 | 0.000 | 33.082 | 0.000 | 5.693 | 0.106 | **244.64** | **0.024** |

N=700

| | Parameters | Univariate LM Test | | Bootstrap LM Test | | Bootstrap W Test | | LRT | |
|---|---|---|---|---|---|---|---|---|---|
| | | LM test | P-values | LM test | P-values | Chi-square | P-values | Chi-square | P-values |
| 1 | **F2, x6** | 795.318 | 0.000 | 766.975 | 0.000 | **437.878** | **0.000** | **794.400** | **0.000** |
| 2 | **F1, x9** | 215.067 | 0.000 | 214.862 | 0.000 | **183.251** | **0.000** | **561.740** | **0.000** |
| 3 | **F2, x20** | 203.848 | 0.000 | 201.634 | 0.000 | **187.452** | **0.000** | **321.070** | **0.000** |
| 4 | F3, x6 | 128.206 | 0.000 | 135.198 | 0.000 | 1.177 | 0.484 | 319.520 | 0.214 |
| 5 | F1, x20 | 108.18 | 0.000 | 105.039 | 0.000 | 1.102 | 0.470 | 319.460 | 0.800 |
| 6 | F2, x8 | 82.385 | 0.000 | 79.944 | 0.000 | 3.063 | 0.240 | **315.180** | **0.039** |
| 7 | **F3, x16** | 77.868 | 0.000 | 78.092 | 0.000 | **60.078** | **0.000** | **240.780** | **0.000** |
| 8 | x6, x20 | 67.495 | 0.000 | 68.539 | 0.000 | 2.224 | 0.326 | 239.580 | 0.273 |
| 9 | F2, x4 | 59.623 | 0.000 | 58.419 | 0.000 | 1.181 | 0.454 | 239.500 | 0.787 |
| 10 | F2, x3 | 58.6 | 0.000 | 56.355 | 0.000 | 1.630 | 0.416 | 238.640 | 0.353 |
| 11 | F2, x5 | 57.406 | 0.000 | 55.430 | 0.000 | 1.055 | 0.480 | 238.400 | 0.624 |
| 12 | x6, x8 | 43.807 | 0.000 | 42.612 | 0.000 | 1.829 | 0.401 | 237.680 | 0.397 |

N=1000

| | Parameters | Univariate LM Test LM test | P-values | Bootstrap LM Test LM test | P-values | Bootstrap W Test Chi-square | P-values | Likelihood Ratio Test Chi-square | P-values |
|---|---|---|---|---|---|---|---|---|---|
| 1 | **F2, x6** | 1059.751 | 0.000 | 1013.785 | 0.000 | **622.506** | **0.000** | **1231.510** | **0.000** |
| 2 | **F1, x9** | 551.727 | 0.000 | 541.939 | 0.000 | **312.393** | **0.000** | **752.360** | **0.000** |
| 3 | **F2, x20** | 262.35 | 0.000 | 266.353 | 0.000 | **265.379** | **0.000** | **407.060** | **0.000** |
| 4 | F3, x6 | 130.052 | 0.000 | 140.235 | 0.000 | 1.257 | 0.455 | 404.750 | 0.129 |
| 5 | **F3, x16** | 128.449 | 0.000 | 128.355 | 0.000 | **97.400** | **0.000** | **266.690** | **0.000** |
| 6 | F2, x8 | 101.117 | 0.000 | 96.034 | 0.000 | 1.369 | 0.460 | 266.590 | 0.761 |
| 7 | F1, x20 | 95.897 | 0.000 | 91.887 | 0.000 | 1.125 | 0.477 | 266.590 | 0.929 |
| 8 | x6, x20 | 79.496 | 0.000 | 83.095 | 0.000 | 3.020 | 0.270 | 264.710 | 0.171 |
| 9 | F2, x5 | 78.413 | 0.000 | 74.127 | 0.000 | 1.188 | 0.471 | 264.710 | 0.940 |
| 10 | F2, x3 | 76.011 | 0.000 | 74.053 | 0.000 | 1.042 | 0.484 | 264.330 | 0.537 |
| 11 | F2. x4 | 74.874 | 0.000 | 72.359 | 0.000 | 1.862 | 0.402 | 263.740 | 0.445 |
| 12 | F2, x1 | 73.657 | 0.000 | 70.275 | 0.000 | 1.403 | 0.447 | 263.600 | 0.702 |

N=2000

| | Parameters | Univariate LM Test LM test | P-values | Bootstrap LM Test LM test | P-values | Bootstrap W Test Chi-square | P-values | LRT Chi-square | P-values |
|---|---|---|---|---|---|---|---|---|---|
| 1 | **F2, x6** | 1962.459 | 0.000 | 1952.303 | 0.000 | **1212.599** | **0.000** | **2087.21** | **0.000** |
| 2 | **F1, x9** | 943.124 | 0.000 | 944.285 | 0.000 | **598.461** | **0.000** | **1240.16** | **0.000** |
| 3 | **F2, x20** | 587.028 | 0.000 | 586.605 | 0.000 | **528.742** | **0.000** | **538.86** | **0.000** |
| 4 | F1, x20 | 283.966 | 0.000 | 277.914 | 0.000 | 1.541 | 0.427 | 538.5 | 0.552 |
| 5 | **F3, x16** | 281.457 | 0.000 | 284.288 | 0.000 | **232.153** | **0.000** | **224.93** | **0.000** |
| 6 | F2, x8 | 207.045 | 0.000 | 207.408 | 0.000 | 1.190 | 0.460 | 224.84 | 0.755 |
| 7 | F3, x6 | 187.422 | 0.000 | 198.224 | 0.000 | 1.196 | 0.479 | 224.75 | 0.766 |
| 8 | x6, x20 | 153.37 | 0.000 | 156.145 | 0.000 | 1.003 | 0.495 | 224.65 | 0.751 |
| 9 | F2, x3 | 144.77 | 0.000 | 145.718 | 0.000 | 1.028 | 0.487 | 224.05 | 0.438 |
| 10 | F2, x5 | 140.543 | 0.000 | 141.952 | 0.000 | 1.995 | 0.374 | 224.02 | 0.887 |
| 11 | F2, x1 | 116.233 | 0.000 | 116.989 | 0.000 | 0.946 | 0.510 | 222.33 | 0.192 |
| 12 | F2, x4 | 114.175 | 0.000 | 114.271 | 0.000 | 9.140 | 0.059 | **215.66** | **0.010** |

## Robustness of the Improved LM Test

### *Varying Factor Correlations*

To further examine the robustness of the improved LM test in covariance structure, we examine these statistical properties by probing varying factor correlations, indicators per factor, and loadings. First, we begin the test with low factor correlations by setting the factor correlations as: (F1, F2) = 0.13, (F1, F3) = 0.1, and (F2, F3) = 0.18. The factor loadings and the number of factors per factor remain the same as in the original model. We find that when factor correlations are low, the improved LM test becomes more efficient at detecting omitted parameters. The performances of the improved LM test and the likelihood ratio test become similar when the sample sizes exceed 100. However, when sample sizes are smaller than 100, the improved LM test still outperforms the LRT.

For high factor correlations, we set the factor correlations as: (F1, F2) = 0.65, (F1, F3) = 0.7, and (F2, F3) = 0.8. We find that high factor correlations tend to behave differently than low factor correlations. The improved LM test tends to accept false parameters, except when N=100. In other sample sizes in this study, the improved LM test falsely detects one additional omitted parameter. However, the improved LM test still outperforms the LRT, which tends to falsely detect three or four additional omitted parameters. Overall, when the factor correlations are high, it increases the potential relationships among factor loadings and residuals. Still, the improved LM test delivers outstanding performance.

**Low Factor Correlations**

$$\mathbf{\Phi} = \begin{bmatrix} 1 & & \\ 0.13 & 1 & \\ 0.10 & 0.18 & 1 \end{bmatrix}$$

**High Factor Correlations**

$$\mathbf{\Phi} = \begin{bmatrix} 1 & & \\ 0.65 & 1 & \\ 0.70 & 0.80 & 1 \end{bmatrix}$$

## Table A2. Low Factor Correlations

N=100

| | | Univariate LM Test | | Bootstrap LM Test | | Bootstrap W Test | | LRT | |
|---|---|---|---|---|---|---|---|---|---|
| | **Parameters** | LM test | P-values | LM test | P-values | Chi-square | P-values | Chi-square | P-values |
| 1 | **F2, x6** | 60.932 | 0.000 | 53.067 | 0.000 | **50.185** | **0.000** | **362.49** | **0.000** |
| 2 | **F2, x20** | 37.404 | 0.000 | 37.194 | 0.000 | **22.676** | **0.000** | **315.72** | **0.000** |
| 3 | F1, x20 | 29.672 | 0.000 | 28.973 | 0.000 | 0.945 | 0.497 | 315.2 | 0.469 |
| 4 | **F3, x16** | 25.418 | 0.000 | 26.322 | 0.000 | **8.971** | **0.033** | **284.37** | **0.000** |
| 5 | **F1, x9** | 15.863 | 0.000 | 16.266 | 0.025 | **19.313** | **0.001** | **261.69** | **0.000** |
| 6 | x3, x14 | 10.051 | 0.002 | 10.686 | 0.020 | 6.580 | 0.075 | **255.25** | **0.011** |
| 7 | x4, x11 | 8.984 | 0.003 | 9.437 | 0.038 | 3.463 | 0.223 | 251.43 | 0.051 |
| 8 | F2, x4 | 8.875 | 0.003 | 8.211 | 0.044 | 1.274 | 0.454 | 251.01 | 0.515 |
| 9 | x6, x11 | 8.064 | 0.005 | 8.824 | 0.044 | 2.739 | 0.273 | 249.31 | 0.192 |
| 10 | F3, x1 | 7.305 | 0.007 | 7.725 | 0.067 | | | **243.89** | **0.020** |
| 11 | F2, x24 | 6.827 | 0.009 | 7.492 | 0.048 | 4.269 | 0.140 | 240.16 | 0.054 |
| 12 | x5, x11 | 6.704 | 0.010 | 7.052 | 0.059 | | | 238.44 | 0.190 |

N=150

| | | Univariate LM Test | | Bootstrap LM Test | | Bootstrap W Test | | LRT | |
|---|---|---|---|---|---|---|---|---|---|
| | Parameters | LM test | P-values | LM test | P-values | Chi-square | P-values | Chi-square | P-values |
| 1 | **F2, x6** | 93.549 | 0.000 | 96.666 | 0.000 | **102.587** | **0.000** | **431.09** | **0.000** |
| 2 | **F1, x9** | 74.239 | 0.000 | 74.825 | 0.000 | **45.244** | **0.000** | **351.33** | **0.000** |
| 3 | **F2, x20** | 46.276 | 0.000 | 44.563 | 0.000 | **53.075** | **0.000** | **296.64** | **0.000** |
| 4 | **F3, x16** | 26.57 | 0.000 | 27.294 | 0.000 | **23.957** | **0.000** | **268.17** | **0.000** |
| 5 | x6, x20 | 25.647 | 0.000 | 22.951 | 0.003 | 1.253 | 0.471 | 268.13 | 0.845 |
| 6 | F3, x6 | 17.288 | 0.000 | 18.813 | 0.038 | 1.126 | 0.465 | 267.95 | 0.671 |
| 7 | F1, x12 | 15.881 | 0.000 | 16.930 | 0.004 | 2.049 | 0.379 | 266.67 | 0.258 |
| 8 | F2, x3 | 13.088 | 0.000 | 13.248 | 0.005 | 2.476 | 0.295 | 264.46 | 0.137 |
| 9 | x12, x3 | 8.544 | 0.003 | 10.035 | 0.039 | 3.412 | 0.231 | 261.58 | 0.090 |
| 10 | F2, x5 | 7.851 | 0.005 | 9.083 | 0.034 | 1.498 | 0.415 | 261.11 | 0.495 |
| 11 | F1, x14 | 7.55 | 0.006 | 8.292 | 0.063 | | | 259.16 | 0.163 |
| 12 | x20, x1 | 7.167 | 0.007 | 8.374 | 0.051 | | | 256.83 | 0.126 |

N=200

| | | Univariate LM Test | | Bootstrap LM Test | | Bootstrap W Test | | LRT | |
|---|---|---|---|---|---|---|---|---|---|
| | Parameters | LM test | P-values | LM test | P-values | Chi-square | P-values | Chi-square | P-values |
| 1 | **F2, x6** | 129.148 | 0.000 | 132.049 | 0.000 | **189.628** | **0.000** | **527.73** | **0.000** |
| 2 | **F1, x9** | 96.736 | 0.000 | 96.592 | 0.000 | **60.397** | **0.000** | **409.11** | **0.000** |
| 3 | **F2, x20** | 95.234 | 0.000 | 92.628 | 0.000 | **107.738** | **0.000** | **283.53** | **0.000** |
| 4 | x6, x20 | 60.975 | 0.000 | 58.134 | 0.000 | 5.720 | 0.122 | 281.63 | 0.169 |
| 5 | **F3, x16** | 56.573 | 0.000 | 56.833 | 0.000 | **29.174** | **0.000** | **215.65** | **0.000** |
| 6 | F2, x5 | 12.703 | 0.000 | 14.346 | 0.009 | 2.192 | 0.322 | 214.51 | 0.284 |
| 7 | F1, x14 | 9.528 | 0.002 | 10.839 | 0.025 | 5.400 | 0.158 | 211.14 | 0.067 |
| 8 | x6, x9 | 9.174 | 0.002 | 9.715 | 0.024 | 1.025 | 0.483 | 211.14 | 0.969 |
| 9 | F1, x20 | 9.087 | 0.003 | 12.098 | 0.030 | 1.787 | 0.403 | 209.59 | 0.213 |
| 10 | F1, x13 | 8.388 | 0.004 | 8.950 | 0.035 | 2.504 | 0.285 | 208.29 | 0.254 |
| 11 | F1, x12 | 8.354 | 0.004 | 9.662 | 0.036 | 1.721 | 0.401 | 207.58 | 0.399 |
| 12 | x20, x3 | 8.052 | 0.005 | 9.536 | 0.041 | 5.016 | 0.157 | 203.99 | 0.058 |

N=500

| | Parameters | Univariate LM Test | | Bootstrap LM Test | | Bootstrap W Test | | LRT | |
|---|---|---|---|---|---|---|---|---|---|
| | | LM test | P-values | LM test | P-values | Chi-square | P-values | Chi-square | P-values |
| 1 | **F2, x6** | 485.053 | 0.000 | 421.198 | 0.000 | **225.103** | **0.000** | **800.350** | **0.000** |
| 2 | **F1, x9** | 265.282 | 0.000 | 242.423 | 0.000 | **142.420** | **0.000** | **583.370** | **0.000** |
| 3 | **F2, x20** | 191.987 | 0.000 | 193.447 | 0.000 | **123.820** | **0.000** | **322.420** | **0.000** |
| 4 | F1, x20 | 97.899 | 0.000 | 86.324 | 0.000 | 0.954 | 0.507 | 322.420 | 0.966 |
| 5 | **F3, x16** | 84.702 | 0.000 | 83.921 | 0.000 | **58.086** | **0.000** | **232.880** | **0.000** |
| 6 | x6, x20 | 58.97 | 0.000 | 66.617 | 0.000 | 1.606 | 0.413 | 232.370 | 0.478 |
| 7 | F2, x8 | 49.468 | 0.000 | 40.816 | 0.000 | 1.173 | 0.491 | 231.900 | 0.493 |
| 8 | F2, x4 | 46.935 | 0.000 | 38.490 | 0.000 | 0.781 | 0.549 | 231.900 | 0.973 |
| 9 | F2, x3 | 42.394 | 0.000 | 34.504 | 0.000 | 0.731 | 0.551 | 231.840 | 0.804 |
| 10 | F2, x5 | 40.422 | 0.000 | 32.449 | 0.000 | 1.027 | 0.492 | 231.830 | 0.940 |
| 11 | F2, x1 | 34.25 | 0.000 | 28.318 | 0.000 | 0.763 | 0.536 | 230.700 | 0.288 |
| 12 | F2, x2 | 29.197 | 0.000 | 23.038 | 0.003 | 1.724 | 0.426 | 230.280 | 0.516 |

N=1000

| | Parameters | Univariate LM Test | | Bootstrap LM Test | | Bootstrap W Test | | LRT | |
|---|---|---|---|---|---|---|---|---|---|
| | | LM test | P-values | LM test | P-values | Chi-square | P-values | Chi-square | P-values |
| 1 | **F2, x6** | 708.561 | 0.000 | 710.403 | 0.000 | **846.215** | **0.000** | **1544.45** | **0.000** |
| 2 | **F1, x9** | 486.137 | 0.000 | 490.303 | 0.000 | **315.185** | **0.000** | **982.67** | **0.000** |
| 3 | **F2, x20** | 429.001 | 0.000 | 428.120 | 0.000 | **422.994** | **0.000** | **403.84** | **0.000** |
| 4 | x6, x20 | 317.4 | 0.000 | 314.413 | 0.000 | 3.245 | 0.263 | 402.27 | 0.211 |
| 5 | **F3, x16** | 131.355 | 0.000 | 130.182 | 0.000 | **100.472** | **0.000** | **250.34** | **0.000** |
| 6 | F3, x6 | 71.884 | 0.000 | 72.057 | 0.000 | 1.981 | 0.379 | 249.26 | 0.298 |
| 7 | F2, x3 | 49.915 | 0.000 | 50.732 | 0.000 | 2.555 | 0.313 | 247.56 | 0.192 |
| 8 | F2, x5 | 44.363 | 0.000 | 46.577 | 0.000 | 1.390 | 0.419 | 247 | 0.454 |
| 9 | F2, x8 | 41.153 | 0.000 | 43.445 | 0.000 | 1.103 | 0.490 | 246.65 | 0.553 |
| 10 | F2, x4 | 40.881 | 0.000 | 42.976 | 0.000 | 1.213 | 0.481 | 246.42 | 0.635 |
| 11 | F1, x20 | 37.506 | 0.000 | 40.830 | 0.000 | 1.102 | 0.486 | 246.41 | 0.923 |
| 12 | F1, x11 | 36.728 | 0.000 | 37.736 | 0.000 | 3.155 | 0.274 | 244.41 | 0.157 |

# Table A3. High Factor Correlations

N=100

| | | Univariate LM Test | | Bootstrap LM Test | | Bootstrap W Test | | LRT | |
|---|---|---|---|---|---|---|---|---|---|
| | Parameters | LM test | P-values | LM test | P-values | Chi-square | P-values | Chi-square | P-values |
| 1 | **F2, x6** | 41.907 | 0.000 | 38.962 | 0.001 | **30.010** | **0.002** | **347.060** | **0.000** |
| 2 | **F1, x9** | 26.238 | 0.000 | 25.459 | 0.003 | **32.797** | **0.000** | **319.830** | **0.000** |
| 3 | **F2, x20** | 14.735 | 0.000 | 13.936 | 0.038 | **22.694** | **0.003** | **301.050** | **0.000** |
| 4 | F3, x6 | 29.587 | 0.000 | 30.073 | 0.004 | 1.786 | 0.388 | 301.050 | 0.945 |
| 5 | **F3, x16** | 11.231 | 0.001 | 12.466 | 0.021 | **15.413** | **0.011** | **284.700** | **0.000** |
| 6 | F2, x5 | 0.448 | 0.503 | 1.691 | 0.443 | | | **281.050** | **0.056** |
| 7 | F3, x9 | 6.119 | 0.013 | 7.472 | 0.129 | | | **275.640** | **0.020** |
| 8 | F2, x8 | 0.574 | 0.449 | 2.156 | 0.373 | | | **270.870** | **0.029** |
| 9 | F2, x1 | 5.767 | 0.016 | 6.647 | 0.076 | | | 270.080 | 0.373 |
| 10 | F2, x3 | 5.836 | 0.016 | 6.286 | 0.088 | | | 269.500 | 0.448 |
| 11 | x6, x20 | 5.758 | 0.016 | 6.181 | 0.088 | | | 268.080 | 0.234 |
| 12 | F1, x10 | 2.181 | 0.140 | 3.013 | 0.255 | | | 268.080 | 0.998 |

N=150

| | | Univariate LM Test | | Bootstrap LM Test | | Bootstrap W Test | | LRT | |
|---|---|---|---|---|---|---|---|---|---|
| | Parameters | LM test | P-values | LM test | P-values | Chi-square | P-values | Chi-square | P-values |
| 1 | **F2, x6** | 44.696 | 0.00 | 45.507 | 0.013 | **33.619** | **0.000** | **359.910** | **0.000** |
| 2 | **F1, x9** | 31.586 | 0.00 | 33.163 | 0.001 | **33.781** | **0.000** | **294.690** | **0.000** |
| 3 | **F2, x20** | 21.423 | 0.00 | 21.726 | 0.006 | 7.913 | 0.066 | **277.370** | **0.000** |
| 4 | F1, x20 | 17.806 | 0.00 | 18.604 | 0.015 | 7.493 | 0.125 | **272.530** | **0.028** |
| 5 | F1, x16 | 14.743 | 0.00 | 15.077 | 0.006 | 1.859 | 0.388 | 269.900 | 0.105 |
| 6 | x1, x13 | 14.354 | 0.00 | 14.707 | 0.006 | **8.309** | **0.044** | 269.640 | 0.605 |
| 7 | x16, x4 | 13.305 | 0.00 | 13.429 | 0.007 | 6.308 | 0.091 | 269.640 | 0.998 |
| 8 | F3, x9 | 12.996 | 0.00 | 13.814 | 0.021 | 0.965 | 0.486 | 268.510 | 0.288 |
| 9 | x6, x15 | 12.752 | 0.00 | 13.352 | 0.014 | 7.785 | 0.067 | 268.060 | 0.504 |
| 10 | x16, x11 | 12.377 | 0.00 | 13.196 | 0.007 | 6.540 | 0.061 | **263.400** | **0.031** |
| 11 | **F3, x16** | 12.288 | 0.00 | 12.728 | 0.017 | 9.213 | 0.062 | **247.210** | **0.000** |
| 12 | x4, x5 | 11.668 | 0.00 | 11.859 | 0.014 | 2.787 | 0.266 | 247.200 | 0.912 |

N=200

| | | Univariate LM Test | | Bootstrap LM Test | | Bootstrap W Test | | LRT | |
|---|---|---|---|---|---|---|---|---|---|
| | Parameters | LM test | P-values | LM test | P-values | Chi-square | P-values | Chi-square | P-values |
| 1 | **F2, x6** | 89.11 | 0.000 | 80.174 | 0.000 | **58.207** | **0.000** | **359.910** | **0.000** |
| 2 | **F1, x9** | 64.29 | 0.000 | 61.536 | 0.000 | **43.849** | **0.000** | **294.690** | **0.000** |
| 3 | **F2, x20** | 26.043 | 0.000 | 26.252 | 0.004 | **18.502** | **0.005** | **277.370** | **0.000** |
| 4 | **F3, x16** | 18.347 | 0.000 | 18.584 | 0.002 | **15.146** | **0.003** | **256.890** | **0.000** |
| 5 | x9, x3 | 17.486 | 0.000 | 17.249 | 0.002 | **8.612** | **0.030** | **248.100** | **0.003** |
| 6 | F1, x20 | 14.602 | 0.000 | 16.296 | 0.035 | 4.691 | 0.141 | **242.570** | **0.019** |
| 7 | F3, x6 | 13.922 | 0.000 | 14.909 | 0.044 | 2.571 | 0.322 | 241.620 | 0.328 |
| 8 | x5, x8 | 13.144 | 0.000 | 14.036 | 0.006 | 5.368 | 0.106 | **237.240** | **0.036** |
| 9 | F2, x7 | 12.727 | 0.000 | 13.220 | 0.020 | 1.718 | 0.400 | 236.690 | 0.457 |
| 10 | F1, x13 | 12.502 | 0.000 | 12.203 | 0.012 | 4.092 | 0.186 | 234.100 | 0.108 |
| 11 | F3, x9 | 12.237 | 0.000 | 13.468 | 0.037 | 1.337 | 0.443 | 233.940 | 0.683 |
| 12 | F1, x14 | 11.527 | 0.001 | 12.253 | 0.009 | 2.551 | 0.307 | 232.390 | 0.214 |

N=500

| | | Univariate LM Test | | Bootstrap LM Test | | Bootstrap W Test | | LRT | |
|---|---|---|---|---|---|---|---|---|---|
| | Parameters | LM test | P-values | LM test | P-values | Chi-square | P-values | Chi-square | P-values |
| 1 | **F2, x6** | 164.205 | 0 | 161.801 | 0.000 | **106.944** | **0.000** | **446.700** | **0.000** |
| 2 | **F1, x9** | 88.361 | 0 | 87.683 | 0.000 | **98.705** | **0.000** | **344.040** | **0.000** |
| 3 | **F3, x16** | 32.303 | 0 | 33.085 | 0.000 | **19.113** | **0.003** | **319.680** | **0.000** |
| 4 | **F2, x20** | 32.067 | 0 | 31.884 | 0.000 | **31.672** | **0.000** | **278.750** | **0.000** |
| 5 | F3, x6 | 31.265 | 0 | 32.651 | 0.000 | 1.565 | 0.400 | 278.660 | 0.764 |
| 6 | F3, x9 | 29.569 | 0 | 30.356 | 0.000 | 1.795 | 0.359 | 277.800 | 0.356 |
| 7 | x20, x19 | 20.414 | 0 | 20.591 | 0.001 | **11.887** | **0.017** | **267.320** | **0.001** |
| 8 | x8, x7 | 19.054 | 0 | 19.405 | 0.004 | 6.634 | 0.081 | **260.960** | **0.012** |
| 9 | F2, x8 | 15.326 | 0 | 16.054 | 0.005 | 1.559 | 0.410 | 260.180 | 0.377 |
| 10 | x5, x13 | 15.042 | 0 | 15.696 | 0.007 | 8.035 | 0.054 | **252.600** | **0.006** |
| 11 | x8, x3 | 14.733 | 0 | 15.831 | 0.004 | 2.603 | 0.281 | 250.900 | 0.192 |
| 12 | x4, x14 | 14.435 | 0 | 15.218 | 0.007 | 7.659 | 0.058 | **243.930** | **0.008** |

N=1000

| | | Univariate LM Test | | Bootstrap LM Test | | Bootstrap W Test | | LRT | |
|---|---|---|---|---|---|---|---|---|---|
| | Parameters | LM test | P-values | LM test | P-values | Chi-square | P-values | Chi-square | P-values |
| 1 | **F2, x6** | 366.254 | 0.000 | 182.761 | 0.000 | **180.103** | **0.000** | **596.550** | **0.000** |
| 2 | **F1, x9** | 182.128 | 0.000 | 118.322 | 0.000 | **150.891** | **0.000** | **375.770** | **0.000** |
| 3 | **F3, x6** | 117.022 | 0.000 | 65.903 | 0.000 | 0.966 | 0.497 | 373.910 | 0.173 |
| 4 | F3, x9 | 64.765 | 0.000 | 58.196 | 0.000 | 0.982 | 0.516 | 372.500 | 0.234 |
| 5 | F2, x20 | 58.591 | 0.000 | 55.265 | 0.000 | **71.717** | **0.000** | **298.200** | **0.000** |
| 6 | **F3, x16** | 54.699 | 0.000 | 55.265 | 0.000 | **33.139** | **0.000** | **257.260** | **0.000** |
| 7 | x4, x5 | 33.114 | 0.000 | 34.254 | 0.000 | 4.192 | 0.201 | 254.220 | 0.081 |
| 8 | F2, x8 | 29.85 | 0.000 | 29.849 | 0.000 | 1.190 | 0.482 | 253.930 | 0.592 |
| 9 | x5, x12 | 29.526 | 0.000 | 29.757 | 0.000 | 3.728 | 0.211 | 250.970 | 0.085 |
| 10 | x10, x12 | 27.98 | 0.000 | 29.342 | 0.000 | **8.361** | **0.044** | **243.270** | **0.006** |
| 11 | F1, x16 | 27.139 | 0.000 | 28.064 | 0.001 | 1.462 | 0.446 | 242.710 | 0.454 |
| 12 | F2, x1 | 25.286 | 0.000 | 26.123 | 0.000 | 2.019 | 0.362 | 241.690 | 0.313 |

## *Varying Number of Indicators Per Factor*

To explore the performance of the improved LM test with different numbers of indicators per factor, we choose to test 5 and 12 loadings per factor.

### 5-Indicators Per Factor

For the 5-indicator model, we have the following factor loadings and factor correlation matrix:

$$\boldsymbol{\Lambda}' = \begin{bmatrix} 0.5, 0.424, 0.581, 0.48, 0.55, 0, 0, 0, 0, 0, 0, 0, 0, 0, 0 \\ 0, 0.65, 0, 0, 0, 0.55, 0.38, 0.6, 0.85, 0.6, 0, 0, 0, 0, 0 \\ 0, 0, 0, 0, 0, 0, 0, 0, 0, 0.65, 0.58, 0.62, 0.49, 0.62, 0.64 \end{bmatrix}$$

$$\mathbf{\Phi} = \begin{bmatrix} 1 & & \\ 0.173 & 1 & \\ 0.408 & 0.262 & 1 \end{bmatrix}$$

## 12-Indicators Per Factor

To test how 12 indicators per factor affect the performance of the improved LM test, we use the following factor loading and factor correlation design:

$$\mathbf{\Lambda}' =$$

$$\begin{bmatrix} 0.65, 0.65, 0.7, 0.7, 0.7, 0.7,0.6, 0.5,0.5, 0.5, 0.6, 0.55, 0.5, 0, 0, 0, 0, 0, 0, 0, 0, 0, 0, 0, 0, 0, 0, 0, 0, 0, 0, 0, 0, 0, 0, 0 \\ 0, 0, 0, 0, 0, 0.5, 0, 0, 0, 0, 0, 0, 0.7, 0.5, 0.5, 0.65,0.5, 0.5,0.6, 0.55, 0.6, 0.45, 0.5, 0.45, 0, 0, 0, 0.5, 0, 0, 0, 0, 0, 0, 0, 0 \\ 0, 0, 0, 0, 0, 0, 0, 0, 0, 0, 0, 0, 0, 0, 0, 0.45, 0, 0, 0, 0, 0, 0, 0, 0, 0.5, 0.5, 0.5, 0.6, 0.6, 0.6, 0.7, 0.70, 0.450, 0.5,0.65,0.55 \end{bmatrix}$$

$$\mathbf{\Phi} = \begin{bmatrix} 1 & & \\ 0.3 & 1 & \\ 0.4 & 0.5 & 1 \end{bmatrix}$$

To ensure the covariance matrix is positive definite, we modify the factor loadings and factor correlations for both the 5-indicator and 12-indicator models. In the 5-indicator model, high factor correlations can result in a covariance matrix that is not positive definite, causing convergence issues, particularly with small sample sizes. To address this, we lower the factor correlations in the 5-indicator model, enabling the model to run successfully. Additionally, we set the population model with two additional parameters: (F2, x2) and (F3, x10). For the 12-indicator model, we set the population model with four additional parameters: (F2, x28), (F1, x13), (F3, x16), and (F2, x6). These parameters are omitted in their respective analysis models. Therefore, we expect the improved LM test to detect these omitted parameters correspondingly.

The results show consistent patterns. As Table A4 shows, when the number of indicators is 5, the statistical power of the improved LM test weakens with smaller sample sizes, though it still performs better than the LRT. In contrast, Table A5 shows that with 12 indicators, both the improved LM tests and the LRT demonstrate equivalent performance. However, when N=300, the improved LM test outperforms the LRT.

Table A4. 5-Indicators Per Factor

N=100

| | | Univariate LM Test | | Bootstrap LM Test | | Bootstrap W Test | | LRT | |
|---|---|---|---|---|---|---|---|---|---|
| | Parameters | LM test | P-values | LM test | P-values | Chi-square | P-values | Chi-square | P-values |
| 1 | **F2, x2** | 42.231 | 0.000 | 35.039 | 0.019 | **13.816** | **0.011** | **172.143** | **0.000** |
| 2 | **F3, x10** | 32.355 | 0.000 | 31.235 | 0.000 | **25.231** | **0.000** | **92.602** | **0.000** |
| 3 | F3, x9 | 13.890 | 0.000 | 13.640 | 0.005 | 2.058 | 0.379 | 89.901 | 0.100 |
| 4 | x4, x5 | 12.794 | 0.000 | 13.206 | 0.033 | 2.136 | 0.332 | 89.611 | 0.590 |
| 5 | F1, x10 | 12.304 | 0.000 | 12.644 | 0.068 | | | 87.136 | 0.116 |
| 6 | F2, x4 | 10.209 | 0.001 | 8.195 | 0.038 | 1.881 | 0.374 | 87.016 | 0.729 |
| 7 | x10, x7 | 7.410 | 0.006 | 8.106 | 0.038 | 1.326 | 0.449 | 86.91 | 0.745 |
| 8 | x4, x3 | 6.002 | 0.014 | 7.446 | 0.097 | | | **82.625** | **0.038** |
| 9 | x10, x11 | 5.829 | 0.016 | 6.703 | 0.075 | | | 82.533 | 0.762 |
| 10 | x9, x7 | 5.766 | 0.016 | 6.923 | 0.095 | | | 82.52 | 0.908 |
| 11 | x10, x15 | 5.056 | 0.025 | 7.901 | 0.051 | | | 79.028 | 0.062 |
| 12 | x6, x13 | 4.601 | 0.032 | 3.669 | 0.237 | | | 78.537 | 0.484 |

N=200

| | | Univariate LM Test | | Bootstrap LM Test | | Bootstrap W Test | | LRT | |
|---|---|---|---|---|---|---|---|---|---|
| | Parameters | LM test | P-values | LM test | P-values | Chi-square | P-values | Chi-square | P-values |
| 1 | **F3, x10** | 70.115 | 0.000 | 64.942 | 0.000 | **38.960** | **0.000** | **127.321** | **0.000** |
| 2 | **F2, x2** | 26.225 | 0.000 | 21.693 | 0.027 | **20.182** | **0.002** | **92.602** | **0.000** |
| 3 | F3, x9 | 19.891 | 0.000 | 17.334 | 0.002 | 3.066 | 0.217 | 89.901 | 0.100 |
| 4 | x2, x9 | 11.820 | 0.001 | 13.475 | 0.023 | 1.435 | 0.437 | 89.577 | 0.569 |
| 5 | x10, x14 | 9.485 | 0.002 | 8.242 | 0.058 | | | 88.837 | 0.389 |
| 6 | x10, x9 | 7.768 | 0.005 | 6.570 | 0.063 | | | 88.660 | 0.674 |
| 7 | F1, x10 | 7.695 | 0.006 | 6.737 | 0.181 | | | 86.759 | 0.168 |
| 8 | x9, x6 | 6.497 | 0.011 | 11.099 | 0.089 | | | 86.507 | 0.616 |
| 9 | x10, x12 | 6.468 | 0.011 | 6.135 | 0.095 | | | 84.850 | 0.198 |
| 10 | x2, x7 | 5.898 | 0.015 | 7.687 | 0.063 | | | 81.974 | 0.090 |
| 11 | x3, x4 | 5.878 | 0.015 | 7.121 | 0.087 | | | **76.803** | **0.023** |
| 12 | x7, x15 | 5.050 | 0.025 | 5.870 | 0.115 | | | 73.151 | 0.056 |

N=300

| | Parameters | Univariate LM Test LM test | P-values | Bootstrap LM Test LM test | P-values | Bootstrap W Test Chi-square | P-values | LRT Chi-square | P-values |
|---|---|---|---|---|---|---|---|---|---|
| 1 | **F3, x10** | 90.856 | 0.000 | 90.551 | 0.000 | **40.332** | **0.000** | **148.301** | **0.000** |
| 2 | **F2, x2** | 58.402 | 0.000 | 57.441 | 0.000 | **42.434** | **0.000** | **87.091** | **0.000** |
| 3 | F3, x9 | 16.693 | 0.000 | 16.033 | 0.006 | 3.055 | 0.250 | 85.820 | 0.260 |
| 4 | F3, x8 | 14.629 | 0.000 | 15.468 | 0.005 | 4.003 | 0.160 | 82.153 | 0.055 |
| 5 | F3, x2 | 11.860 | 0.001 | 12.243 | 0.028 | 0.916 | 0.514 | 82.151 | 0.971 |
| 6 | F2, x4 | 11.205 | 0.001 | 11.540 | 0.020 | 4.329 | 0.159 | 78.433 | 0.054 |
| 7 | x10, x11 | 11.120 | 0.001 | 12.126 | 0.017 | 3.050 | 0.245 | 76.721 | 0.191 |
| 8 | x10, x8 | 9.581 | 0.002 | 10.490 | 0.021 | 2.024 | 0.359 | 75.720 | 0.317 |
| 9 | x7, x15 | 8.644 | 0.003 | 9.737 | 0.033 | 7.667 | 0.058 | **68.972** | **0.009** |
| 10 | F3, x6 | 7.044 | 0.008 | 8.681 | 0.067 | | | 68.863 | 0.742 |
| 11 | x10, x15 | 7.005 | 0.008 | 7.098 | 0.052 | | | 68.856 | 0.935 |
| 12 | x2, x9 | 6.984 | 0.008 | 7.794 | 0.050 | | | 68.010 | 0.357 |

N=500

| | Parameters | Univariate LM Test LM test | P-values | Bootstrap LM Test LM test | P-values | Bootstrap W Test Chi-square | P-values | LRT Chi-square | P-values |
|---|---|---|---|---|---|---|---|---|---|
| 1 | **F3, 10** | 201.014 | 0.000 | 198.204 | 0.000 | **72.095** | **0.000** | **219.622** | **0.000** |
| 2 | **F2, x2** | 136.377 | 0.000 | 131.367 | 0.000 | **49.903** | **0.000** | **79.244** | **0.000** |
| 3 | F1, x10 | 38.392 | 0.000 | 33.852 | 0.019 | 1.035 | 0.484 | 78.724 | 0.471 |
| 4 | F3, x9 | 33.370 | 0.000 | 32.237 | 0.000 | 1.121 | 0.481 | 78.698 | 0.871 |
| 5 | F3, x6 | 30.226 | 0.000 | 29.281 | 0.000 | 1.309 | 0.458 | 78.698 | 0.991 |
| 6 | F3, x8 | 20.525 | 0.000 | 19.937 | 0.003 | 1.778 | 0.396 | 78.695 | 0.958 |
| 7 | x10, x6 | 20.465 | 0.000 | 20.237 | 0.002 | 1.036 | 0.484 | 77.856 | 0.360 |
| 8 | x9, x6 | 16.426 | 0.000 | 21.456 | 0.023 | 1.156 | 0.468 | 77.728 | 0.720 |
| 9 | x10, x11 | 15.756 | 0.000 | 15.514 | 0.004 | 3.013 | 0.235 | 75.76 | 0.161 |
| 10 | x2, x9 | 15.731 | 0.000 | 17.283 | 0.004 | 1.845 | 0.381 | 75.672 | 0.767 |
| 11 | x8, x14 | 14.215 | 0.000 | 14.970 | 0.008 | 8.407 | 0.057 | **68.844** | **0.009** |
| 12 | x6, x8 | 13.067 | 0.000 | 17.412 | 0.032 | 2.118 | 0.375 | 68.044 | 0.371 |

Table A5. 12-Indicators Per Factor

N=100

| | Parameters | Univariate LM Test | | Bootstrap LM Test | | Bootstrap W Test | | LRT | |
|---|---|---|---|---|---|---|---|---|---|
| | | LM test | P-values | LM test | P-values | Chi-square | P-values | Chi-square | P-values |
| 1 | **F2, x28** | 43.572 | 0.000 | 38.423 | 0.002 | **89.339** | **0.000** | **969.88** | **0.000** |
| 2 | **F1, x13** | 43.281 | 0.000 | 42.247 | 0.011 | **24.883** | **0.000** | **829.76** | **0.000** |
| 3 | **F3, x16** | 39.896 | 0.000 | 38.511 | 0.004 | **17.015** | **0.003** | **748.31** | **0.000** |
| 4 | F1, x28 | 13.527 | 0.000 | 11.627 | 0.073 | | | 747.63 | 0.410 |
| 5 | F3, x13 | 19.828 | 0.000 | 21.378 | 0.010 | 1.125 | 0.476 | 746.49 | 0.285 |
| 6 | F1, x16 | 8.008 | 0.005 | 10.433 | 0.059 | | | 746.25 | 0.631 |
| 7 | x28, x13 | 2.385 | 0.123 | 2.575 | 0.273 | | | 746.25 | 0.987 |
| 8 | x13, x16 | 5.367 | 0.021 | 5.669 | 0.074 | | | 745.24 | 0.315 |
| 9 | x2, x12 | 3.379 | 0.066 | 4.384 | 0.167 | | | 744.96 | 0.596 |
| 10 | **F2, x6** | 97.471 | 0.000 | 79.725 | 0.003 | **218.927** | **0.000** | **580.29** | **0.000** |
| 11 | x3, x4 | 0.85 | 0.357 | 2.330 | 0.348 | | | 580.18 | 0.746 |
| 12 | x28, x35 | 3.275 | 0.070 | 3.809 | 0.166 | | | 579.89 | 0.586 |

N=200

| | Parameters | Univariate LM Test | | Bootstrap LM Test | | Bootstrap W Test | | LRT | |
|---|---|---|---|---|---|---|---|---|---|
| | | LM test | P-values | LM test | P-values | Chi-square | P-values | Chi-square | P-values |
| 1 | **F2, x6** | 228.34 | 0.00 | 190.742 | 0.000 | **226.139** | **0.000** | **928.07** | **0.000** |
| 2 | **F2, x28** | 132.093 | 0.00 | 133.615 | 0.000 | **146.055** | **0.000** | **813.04** | **0.000** |
| 3 | **F1, x13** | 131.812 | 0.00 | 119.635 | 0.000 | **77.737** | **0.000** | **675.77** | **0.000** |
| 4 | **F3, x16** | 97.05 | 0.00 | 86.211 | 0.000 | **30.897** | **0.000** | **582.36** | **0.000** |
| 5 | F3, x6 | 77.574 | 0.00 | 72.984 | 0.004 | 1.082 | 0.490 | 581.73 | 0.428 |
| 6 | F3, x13 | 46.315 | 0.00 | 35.594 | 0.007 | 1.444 | 0.413 | 581.38 | 0.556 |
| 7 | F1, x16 | 41.899 | 0.00 | 33.670 | 0.002 | 1.830 | 0.382 | 581.33 | 0.808 |
| 8 | x13, x16 | 33.522 | 0.00 | 26.670 | 0.000 | 1.490 | 0.447 | 581.28 | 0.830 |
| 9 | x28, x16 | 24.574 | 0.00 | 22.794 | 0.001 | 3.353 | 0.197 | 578.19 | 0.079 |
| 10 | x6, x1 | 19.803 | 0.00 | 17.102 | 0.001 | 2.354 | 0.322 | 576.98 | 0.272 |
| 11 | x6, x28 | 19.463 | 0.00 | 19.543 | 0.009 | 2.749 | 0.295 | 575.61 | 0.242 |
| 12 | F2, x32 | 19.428 | 0.00 | 19.573 | 0.003 | 3.784 | 0.217 | 573.02 | 0.107 |

N=300

| | Parameters | Univariate LM Test | | Bootstrap LM Test | | Bootstrap W Test | | LRT | |
|---|---|---|---|---|---|---|---|---|---|
| | | LM test | P-values | LM test | P-values | Chi-square | P-values | Chi-square | P-values |
| 1 | **F2, x6** | 221.084 | 0.000 | 222.643 | 0.000 | **373.677** | **0.000** | **1252.080** | **0.000** |
| 2 | **F1, x13** | 202.705 | 0.000 | 210.428 | 0.000 | **129.587** | **0.000** | **1003.840** | **0.000** |
| 3 | **F2, x28** | 183.313 | 0.000 | 167.474 | 0.000 | **188.098** | **0.000** | **847.300** | **0.000** |
| 4 | **F3, x16** | 130.012 | 0.000 | 129.396 | 0.000 | **54.044** | **0.000** | **653.600** | **0.000** |
| 5 | F3, x6 | 122.427 | 0.000 | 108.683 | 0.000 | 2.288 | 0.308 | 652.000 | 0.205 |
| 6 | F1, x16 | 46.094 | 0.000 | 50.496 | 0.000 | 0.882 | 0.523 | 651.930 | 0.783 |
| 7 | x13, x16 | 43.626 | 0.000 | 47.956 | 0.000 | 2.716 | 0.307 | 650.480 | 0.230 |
| 8 | F2, x5 | 25.399 | 0.000 | 26.701 | 0.000 | 2.305 | 0.324 | 649.580 | 0.343 |
| 9 | F2, x32 | 25.394 | 0.000 | 23.674 | 0.002 | 1.740 | 0.378 | 648.930 | 0.421 |
| 10 | F3, x13 | 25.221 | 0.000 | 44.820 | 0.028 | 1.168 | 0.462 | 648.770 | 0.688 |
| 11 | x29, x30 | 23.209 | 0.000 | 23.647 | 0.000 | **8.976** | **0.036** | **639.830** | **0.003** |
| 12 | F2, x4 | 22.214 | 0.000 | 22.793 | 0.001 | 1.959 | 0.373 | 639.110 | 0.398 |

N=500

| | | Univariate LM Test | | Bootstrap LM Test | | Bootstrap W Test | | LRT | |
|---|---|---|---|---|---|---|---|---|---|
| | Parameters | LM test | P-values | LM test | P-values | Chi-square | P-values | Chi-square | P-values |
| 1 | **F1, x13** | 461.937 | 0 | 404.435 | 0.000 | **198.847** | **0.000** | **1410.39** | **0.000** |
| 2 | **F2, x6** | 400.032 | 0 | 367.705 | 0.000 | **596.082** | **0.000** | **1142.54** | **0.000** |
| 3 | **F3, x16** | 292.493 | 0 | 264.810 | 0.000 | **94.089** | **0.000** | **875.36** | **0.000** |
| 4 | F3, x13 | 245.116 | 0 | 170.708 | 0.000 | 5.119 | 0.110 | 874.63 | 0.394 |
| 5 | **F2, x28** | 241.162 | 0 | 268.713 | 0.000 | **346.318** | **0.000** | **563.13** | **0.000** |
| 6 | F1, x16 | 187.052 | 0 | 151.074 | 0.000 | 0.964 | 0.509 | 563.12 | 0.956 |
| 7 | x13, x16 | 117.424 | 0 | 91.630 | 0.000 | 3.345 | 0.253 | 560.75 | 0.123 |
| 8 | F3, x6 | 94.49 | 0 | 120.618 | 0.001 | 4.224 | 0.138 | 556.96 | 0.052 |
| 9 | F1, x28 | 77.131 | 0 | 58.237 | 0.017 | 1.014 | 0.485 | 556.95 | 0.933 |
| 10 | F2, x3 | 54.215 | 0 | 47.388 | 0.000 | 2.013 | 0.336 | 555.02 | 0.165 |
| 11 | x16, x3 | 45.495 | 0 | 36.470 | 0.000 | 2.315 | 0.311 | 553.61 | 0.235 |
| 12 | F2, ~x5 | 33.462 | 0 | 27.694 | 0.001 | 1.151 | 0.484 | 553.61 | 0.951 |

N=1000

| | | Univariate LM Test | | Bootstrap LM Test | | Bootstrap W Test | | LRT | |
|---|---|---|---|---|---|---|---|---|---|
| | Parameters | LM test | P-values | LM test | P-values | Chi-square | P-values | Chi-square | P-values |
| 1 | **F2, x6** | 780.100 | 0.000 | 820.228 | 0.000 | **1045.898** | **0.000** | **2459.97** | **0.000** |
| 2 | **F1, x13** | 711.316 | 0.000 | 736.515 | 0.000 | **368.448** | **0.000** | **1540.19** | **0.000** |
| 3 | **F3, x16** | 573.377 | 0.000 | 566.802 | 0.000 | **194.870** | **0.000** | **1066.67** | **0.000** |
| 4 | F3, x6 | 522.391 | 0.000 | 485.847 | 0.000 | 1.822 | 0.380 | 1066.49 | 0.671 |
| 5 | F1, x16 | 336.246 | 0.000 | 340.162 | 0.000 | 1.828 | 0.368 | 1064.19 | 0.129 |
| 6 | **F2, x28** | 332.581 | 0.000 | 322.924 | 0.000 | **595.541** | **0.000** | **573.91** | **0.000** |
| 7 | F3, x13 | 249.036 | 0.000 | 290.799 | 0.000 | 2.412 | 0.331 | 572.66 | 0.265 |
| 8 | x13, x16 | 161.386 | 0.000 | 163.551 | 0.000 | 0.974 | 0.505 | 572.33 | 0.562 |
| 9 | F2, x5 | 58.493 | 0.000 | 62.417 | 0.000 | 4.157 | 0.196 | 569.75 | 0.109 |
| 10 | x6, x16 | 57.389 | 0.000 | 59.037 | 0.000 | 1.132 | 0.485 | 569.61 | 0.708 |
| 11 | F2, x1 | 50.178 | 0.000 | 53.127 | 0.000 | 1.700 | 0.384 | 569.04 | 0.447 |
| 12 | F2, x4 | 49.058 | 0.000 | 53.468 | 0.000 | 1.252 | 0.464 | 568.84 | 0.658 |

## *Varying Magnitudes of Factor Loadings*

To explore the extent to which factor loadings affect the performance of the improve LM test, we create two sets of loadings: Low and high. For the low factor loading model, we have the following factor loadings and factor correlation matrix:

**Low Factor Loadings**

$$\mathbf{\Lambda}' = \begin{bmatrix} 0.3, 0.3, 0.3, 0.3, 0.3, 0.4, 0.4, 0.35, 0.45, 0, 0, 0, 0, 0, 0,0, \ 0, 0, 0, 0, 0, 0, 0, 0 \\ 0, 0, 0, 0, 0, 0.45, 0, 0, \ 0.3, 0.3, 0.3, 0.3, 0.3, 0.3, 0.3, 0.35, 0, 0, 0, 0.55, 0, 0, 0, 0 \\ 0, 0, 0, 0, 0, 0, 0, 0, \ 0, 0, 0, 0, 0, 0, 0.45, \ 0.3, 0.3, 0.3, 0.3, 0.4, 0.4, 0.3, 0.3 \end{bmatrix}$$

$$
\boldsymbol{\Phi} = \begin{bmatrix} 1 & & \\ 0.3 & 1 & \\ 0.4 & 0.5 & 1 \end{bmatrix}
$$

## High Factor Loadings

For the high factor loading model, we have the following factor loadings and factor correlation matrix:

$$
\boldsymbol{\Lambda}' = \begin{bmatrix} 0.8, 0.8, 0.8, 0.8, 0.8, 0.9, 0.9, 0.85, 0.9, 0, 0, 0, 0, 0, 0,0,\ \ 0, 0, 0, 0, 0, 0, 0, 0 \\ 0, 0, 0, 0, 0, 0.9, 0, 0,\ \ 0.8, 0.8, 0.8, 0.8, 0.8, 0.8, 0.8, 0.85, 0, 0, 0, 0.9, 0, 0, 0, 0 \\ 0, 0, 0, 0, 0, 0, 0, 0,\ \ 0, 0, 0, 0, 0, 0, 0, 0.9,\ \ 0.8, 0.8, 0.8, 0.8, 0.9, 0.9, 0.8, 0.8 \end{bmatrix}
$$

$$
\boldsymbol{\Phi} = \begin{bmatrix} 1 & & \\ 0.3 & 1 & \\ 0.4 & 0.5 & 1 \end{bmatrix}
$$

We find that the magnitudes of factor loadings influence the performance of the improved LM test, and this effect is dependent on the sample size. When sample sizes are greater than 400, we find that the improved LM test delivers efficient and robust performance compared to the LRT. However, low factor loadings in smaller sample sizes tend to have stronger impacts on the detection of omitted variables and convergence. We find that when sample sizes are smaller than 400, the models encounter convergence issues, mainly because the covariance matrix of latent variables becomes not positive definite. In contrast, with high factor loadings, both the improved LM test and LRT perform well across all sample sizes in this study. However, the improved LM test consistently demonstrates a statistical edge in detecting correct parameters compared to the LRT.

## Table A6. Low Factor Loadings

N=400

| | | Univariate LM Test | | Bootstrap LM Test | | Bootstrap W Test | | LRT | |
|---|---|---|---|---|---|---|---|---|---|
| | Parameters | LM test | P-values | LM test | P-values | Chi-square | P-values | Chi-square | P-values |
| 1 | **F2, x6** | 20.061 | 0.000 | 25.328 | 0.057 | **16.576** | **0.010** | **333.08** | **0.000** |
| 2 | F3, x6 | 32.47 | 0.000 | 24.622 | 0.050 | 1.204 | 0.436 | 332.69 | 0.530 |
| 3 | x7, x1 | 15.245 | 0.000 | 18.293 | 0.006 | 5.754 | 0.104 | **327.92** | **0.029** |
| 4 | **F1, x9** | 14.886 | 0.000 | 16.746 | 0.117 | **7.199** | **0.038** | **311.32** | **0.000** |
| 5 | F3, x9 | 13.244 | 0.000 | 11.582 | 0.146 | 1.362 | 0.418 | 310.41 | 0.341 |
| 6 | **F3, x16** | 9.764 | 0.002 | 12.658 | 0.137 | **9.927** | **0.022** | **300.03** | **0.001** |
| 7 | F3, x7 | 9.167 | 0.002 | 12.329 | 0.040 | 3.292 | 0.181 | 299.73 | 0.584 |
| 8 | x6, x20 | 9.07 | 0.003 | 13.587 | 0.043 | 1.103 | 0.486 | 296.57 | 0.075 |
| 9 | F2, x7 | 8.635 | 0.003 | 10.449 | 0.062 | 2.868 | 0.265 | 293.94 | 0.105 |
| 10 | F1, x16 | 8.36 | 0.004 | 10.195 | 0.155 | 1.251 | 0.434 | 293.35 | 0.441 |
| 11 | **F2, x20** | 8.289 | 0.004 | 16.150 | 0.121 | **14.331** | **0.016** | **263.93** | **0.000** |
| 12 | x18, x21 | 8.19 | 0.004 | 12.134 | 0.027 | 4.401 | 0.190 | 260.7 | 0.072 |

N=500

| | | Univariate LM Test | | Bootstrap LM Test | | Bootstrap W Test | | LRT | |
|---|---|---|---|---|---|---|---|---|---|
| | Parameters | LM test | P-values | LM test | P-values | Chi-square | P-values | Chi-square | P-values |
| 1 | **F2, x6** | 36.225 | 0.000 | 33.149 | 0.004 | **10.458** | **0.025** | **307.870** | **0.000** |
| 2 | F3, x6 | 35.652 | 0.000 | 33.176 | 0.005 | 1.284 | 0.448 | **307.850** | **0.000** |
| 3 | x6, x20 | 25.177 | 0.000 | 30.737 | 0.002 | 3.394 | 0.263 | **293.750** | **0.000** |
| 4 | **F2, x20** | 24.87 | 0.000 | 22.958 | 0.076 | **10.593** | **0.026** | 278.410 | NC |
| 5 | **F1=~x9** | 15.835 | 0.000 | 20.820 | 0.009 | **9.432** | **0.012** | **241.850** | **0.000** |
| 6 | F1, x20 | 15.793 | 0.000 | 22.626 | 0.010 | 1.105 | 0.462 | 241.800 | 0.825 |
| 7 | x9, x7 | 14.181 | 0.000 | 19.883 | 0.009 | 4.596 | 0.204 | 239.460 | 0.126 |
| 8 | **F3, x16** | 13.875 | 0.000 | 18.163 | 0.056 | **16.633** | **0.003** | **220.670** | **0.000** |
| 9 | F3, x9 | 13.315 | 0.000 | 12.421 | 0.104 | **7.630** | **0.048** | **212.890** | **0.005** |
| 10 | x16, x21 | 11.465 | 0.001 | 16.556 | 0.011 | 5.106 | 0.151 | 209.310 | 0.058 |
| 11 | x20, x9 | 11.054 | 0.001 | 16.865 | 0.010 | 4.287 | 0.187 | **204.950** | **0.037** |
| 12 | x20, x14 | 9.727 | 0.002 | 14.402 | 0.017 | 6.277 | 0.088 | **200.100** | **0.028** |

N=700

| | | Univariate LM Test | | Bootstrap LM Test | | Bootstrap W Test | | LRT | |
|---|---|---|---|---|---|---|---|---|---|
| | Parameters | LM test | P-values | LM test | P-values | Chi-square | P-values | Chi-square | P-values |
| 1 | **F2, x6** | 39.708 | 0.000 | 37.885 | 0.003 | **22.035** | **0.001** | **370.86** | **0.000** |
| 2 | F3, x6 | 33.269 | 0.000 | 32.607 | 0.003 | 1.252 | 0.467 | 370.74 | 0.734 |
| 3 | **F2, x20** | 28.788 | 0.000 | 27.318 | 0.034 | **22.331** | **0.001** | **329.76** | **0.000** |
| 4 | **F1, x9** | 27.376 | 0.000 | 28.850 | 0.003 | **15.658** | **0.001** | **278.03** | **0.000** |
| 5 | **F3, x16** | 23.307 | 0.000 | 23.340 | 0.013 | **19.770** | **0.001** | **246.33** | **0.000** |
| 6 | x9, x8 | 16.132 | 0.000 | 17.350 | 0.003 | 2.553 | 0.316 | 244.49 | 0.174 |
| 7 | x16, x19 | 15.868 | 0.000 | 16.991 | 0.005 | **10.105** | **0.035** | **234.74** | **0.002** |
| 8 | F1, x16 | 15.379 | 0.000 | 16.669 | 0.009 | 1.872 | 0.340 | 233.55 | 0.274 |
| 9 | x1, x7 | 11.175 | 0.001 | 12.115 | 0.020 | 5.614 | 0.112 | **228.61** | **0.026** |
| 10 | F3, x9 | 10.957 | 0.001 | 12.830 | 0.095 | 0.931 | 0.508 | 228.54 | 0.796 |
| 11 | x6, x20 | 10.815 | 0.001 | 12.212 | 0.029 | 1.346 | 0.447 | 228.24 | 0.584 |
| 12 | x14, x21 | 10.441 | 0.001 | 11.363 | 0.019 | 6.216 | 0.093 | **222.72** | **0.019** |

N=1000

| | Parameters | Univariate LM Test LM test | P-values | Bootstrap LM Test LM test | P-values | Bootstrap W Test Chi-square | P-values | LRT Chi-square | P-values |
|---|---|---|---|---|---|---|---|---|---|
| 1 | **F1, x9** | 58.984 | 0.000 | 62.220 | 0.000 | **30.689** | **0.000** | **398.99** | **0.000** |
| 2 | **F3, x16** | 54.975 | 0.000 | 53.113 | 0.004 | **38.293** | **0.000** | **380.21** | **0.000** |
| 3 | **F2, x6** | 54.463 | 0.000 | 54.411 | 0.000 | **22.254** | **0.001** | **316.53** | **0.000** |
| 4 | F3, x6 | 51.42 | 0.000 | 51.119 | 0.000 | 1.750 | 0.419 | 316.13 | 0.527 |
| 5 | **F2, x20** | 38.254 | 0.000 | 36.668 | 0.022 | **23.690** | **0.001** | **237.82** | **0.000** |
| 6 | F1, x16 | 37.387 | 0.000 | 40.281 | 0.000 | 2.574 | 0.281 | 233.98 | 0.050 |
| 7 | F3, x9 | 36.524 | 0.000 | 34.381 | 0.030 | 1.188 | 0.471 | 233.32 | 0.418 |
| 8 | x16, x21 | 22.832 | 0.000 | 26.541 | 0.001 | 8.124 | 0.051 | **225.63** | **0.006** |
| 9 | x6, x20 | 17.209 | 0.000 | 21.863 | 0.005 | 1.636 | 0.419 | 223.74 | 0.168 |
| 10 | x3, x7 | 13.956 | 0.000 | 17.609 | 0.012 | 5.227 | 0.140 | **219.73** | **0.045** |
| 11 | x16, x6 | 12.291 | 0.000 | 15.575 | 0.012 | 2.469 | 0.317 | 217.93 | 0.180 |
| 12 | F3, x1 | 11.998 | 0.001 | 5.561 | 0.011 | 4.092 | 0.172 | **214.05** | **0.049** |

# Table A7. High Factor Loadings

N=100

| | Parameters | Univariate LM Test LM test | P-values | Bootstrap LM Test LM test | P-values | Bootstrap W Test Chi-square | P-values | LRT Chi-square | P-values |
|---|---|---|---|---|---|---|---|---|---|
| 1 | **F3, x16** | 53.23 | 0.000 | 44.991 | 0.004 | **25.018** | **0.000** | **388.74** | **0.000** |
| 2 | **F2, x6** | 41.87 | 0.000 | 36.235 | 0.001 | **28.377** | **0.000** | **351.97** | **0.000** |
| 3 | **F2, x20** | 37.745 | 0.000 | 32.974 | 0.007 | **27.885** | **0.001** | **315.78** | **0.000** |
| 4 | **F1, x9** | 21.268 | 0.000 | 21.743 | 0.004 | **16.710** | **0.002** | **279.31** | **0.000** |
| 5 | F3, x6 | 20.779 | 0.000 | 19.139 | 0.049 | 1.131 | 0.462 | 279.23 | 0.770 |
| 6 | x9, x3 | 14.834 | 0.000 | 14.988 | 0.003 | **8.663** | **0.027** | **270.35** | **0.003** |
| 7 | F2, x21 | 14.5 | 0.000 | 12.896 | 0.018 | 4.001 | 0.157 | 266.94 | 0.065 |
| 8 | F1, x12 | 13.549 | 0.000 | 12.993 | 0.009 | **6.901** | **0.043** | **257.88** | **0.003** |
| 9 | x16, x10 | 12.251 | 0.000 | 11.882 | 0.006 | 6.413 | 0.066 | **252.49** | **0.020** |
| 10 | x20, x12 | 10.961 | 0.001 | 11.073 | 0.012 | 3.627 | 0.186 | 249.39 | 0.078 |
| 11 | x16, x17 | 10.418 | 0.001 | 10.746 | 0.029 | 3.406 | 0.241 | 247.04 | 0.125 |
| 12 | x21, x12 | 9.593 | 0.002 | 9.847 | 0.034 | 1.633 | 0.395 | 246.38 | 0.416 |

N=300

| | Parameters | Univariate LM Test LM test | P-values | Bootstrap LM Test LM test | P-values | Bootstrap W Test Chi-square | P-values | LRT Chi-square | P-values |
|---|---|---|---|---|---|---|---|---|---|
| 1 | **F2, x6** | 275.93 | 0.000 | 276.700 | 0.000 | **562.319** | **0.000** | **817.25** | **0.000** |
| 2 | **F1, x9** | 209.86 | 0.000 | 211.901 | 0.000 | **196.257** | **0.000** | **515.14** | **0.000** |
| 3 | **F2, x20** | 128.525 | 0.000 | 127.522 | 0.000 | **152.197** | **0.000** | **354.73** | **0.000** |
| 4 | **F3, x16** | 106.783 | 0.000 | 105.149 | 0.000 | **111.193** | **0.000** | **205.65** | **0.000** |
| 5 | x6, x20 | 87.282 | 0.000 | 86.555 | 0.000 | 1.052 | 0.496 | 205.53 | 0.734 |
| 6 | F3, x6 | 83.759 | 0.000 | 83.726 | 0.000 | 4.874 | 0.131 | 203.18 | 0.125 |
| 7 | F1, x15 | 20.623 | 0.000 | 21.242 | 0.001 | 1.691 | 0.394 | 202.29 | 0.345 |
| 8 | F2, x8 | 18.284 | 0.000 | 19.519 | 0.003 | 1.001 | 0.500 | 202.23 | 0.804 |
| 9 | x6, x12 | 17.951 | 0.000 | 18.420 | 0.000 | 2.246 | 0.365 | 202.16 | 0.784 |
| 10 | F1, x12 | 15.639 | 0.000 | 16.806 | 0.004 | 0.809 | 0.530 | 201.86 | 0.586 |
| 11 | F3, x12 | 15.25 | 0.000 | 15.818 | 0.005 | 2.232 | 0.346 | 200.66 | 0.273 |
| 12 | x6, x15 | 14.545 | 0.000 | 15.472 | 0.002 | 4.486 | 0.192 | 197.56 | 0.079 |

N=500

| | Parameters | Univariate LM Test | | Bootstrap LM Test | | Bootstrap W Test | | LRT | |
|---|---|---|---|---|---|---|---|---|---|
| | | LM test | P-values | LM test | P-values | Chi-square | P-values | Chi-square | P-values |
| 1 | **F2, x6** | 433.662 | 0.000 | 434.277 | 0.000 | **664.337** | **0.000** | **1277.97** | **0.000** |
| 2 | **F1, x9** | 347.478 | 0.000 | 347.727 | 0.000 | **346.671** | **0.000** | **756.69** | **0.000** |
| 3 | **F2, x20** | 221.892 | 0.000 | 220.501 | 0.000 | **215.782** | **0.000** | **479.7** | **0.000** |
| 4 | F3, x6 | 168.598 | 0.000 | 169.013 | 0.000 | 5.442 | 0.106 | **447.54** | **0.000** |
| 5 | **F3, x16** | 154.575 | 0.000 | 153.615 | 0.000 | **141.152** | **0.000** | **254.11** | **0.000** |
| 6 | x6, x20 | 145.585 | 0.000 | 145.388 | 0.000 | 1.416 | 0.434 | 254.08 | 0.848 |
| 7 | F2, x8 | 43.358 | 0.000 | 44.265 | 0.000 | 1.911 | 0.350 | 252.69 | 0.238 |
| 8 | x6, x12 | 34.384 | 0.000 | 35.055 | 0.000 | 1.897 | 0.362 | 250.94 | 0.186 |
| 9 | x20, x12 | 33.352 | 0.000 | 34.028 | 0.000 | 1.562 | 0.410 | 250.58 | 0.550 |
| 10 | F3, x12 | 29.333 | 0.000 | 29.731 | 0.000 | 3.666 | 0.214 | 247.83 | 0.097 |
| 11 | x20, x11 | 25.946 | 0.000 | 27.273 | 0.000 | 5.513 | 0.119 | **243.57** | **0.039** |
| 12 | x20, x8 | 24.982 | 0.000 | 25.731 | 0.000 | 2.960 | 0.260 | 241.47 | 0.147 |

N=1000

| | Parameters | Univariate LM Test | | Bootstrap LM Test | | Bootstrap W Test | | LRT | |
|---|---|---|---|---|---|---|---|---|---|
| | | LM test | P-values | LM test | P-values | Chi-square | P-values | Chi-square | P-values |
| 1 | **F2, x6** | 911.297 | 0.00 | 912.366 | 0.000 | **1681.605** | **0.000** | **2427.85** | **0.000** |
| 2 | **F1, x9** | 798.267 | 0.00 | 798.635 | 0.000 | **693.119** | **0.000** | **1220.3** | **0.000** |
| 3 | **F2, x20** | 490.452 | 0.00 | 490.115 | 0.000 | **550.420** | **0.000** | **642.49** | **0.000** |
| 4 | **F3, x16** | 284.323 | 0.00 | 282.955 | 0.000 | **276.200** | **0.000** | **255.31** | **0.000** |
| 5 | x6, x20 | 264.367 | 0.00 | 264.714 | 0.000 | 8.478 | 0.057 | **247.11** | **0.004** |
| 6 | F3, x6 | 254.426 | 0.00 | 255.271 | 0.000 | 1.143 | 0.476 | 246.97 | 0.705 |
| 7 | F2, x8 | 89.066 | 0.00 | 91.158 | 0.000 | 3.188 | 0.254 | 244.94 | 0.154 |
| 8 | x6, x12 | 53.928 | 0.00 | 54.597 | 0.000 | 1.479 | 0.419 | 244.56 | 0.538 |
| 9 | x6, x11 | 50.576 | 0.00 | 50.717 | 0.000 | 1.433 | 0.429 | 244.17 | 0.531 |
| 10 | x9, x8 | 47.468 | 0.00 | 48.918 | 0.000 | 3.124 | 0.264 | 242.05 | 0.145 |
| 11 | F1, x11 | 42.107 | 0.00 | 42.928 | 0.000 | 1.502 | 0.432 | 241.69 | 0.553 |
| 12 | F1, x12 | 40.789 | 0.00 | 41.050 | 0.000 | 1.522 | 0.422 | 241.18 | 0.475 |

# Huddy and Khatib’s (2007) survey questions on the 2002 student sample.

Q20. How similar do you feel to the average American?

1. Very similar
2. Somewhat similar
3. Not very similar
4. Not at all

Q22. When you hear a non-American criticizing Americans, to what extent do you feel you are being personally criticized?

1. A great deal
2. Somewhat
3. Very little

4. Not at all

Q23. How well does the term American describe you?

1. Very well
2. Somewhat well
3. Not very well
4. Not at all

Q24. When talking about Americans, how often would you say "we" rather than "they"

1. Most of the time
2. Some of the time
3. Occasionally
4. Never

Q26. How good does it make you feel when you see the American flag flying:?

1. Extremely good
2. Very good
3. Somewhat good
4. Or not very good

Q27. How angry does it make you feel, if at all, when you hear someone criticizing the United States:

1. Extremely angry
2. Somewhat angry
3. Not very angry
4. Not at all angry

Q28. How proud do you feel when you hear the national anthem?

1. Extremely proud
2. Very proud
3. Somewhat proud
4. Or not very proud

Q37. How strongly do you agree or disagree with the following statements? There is too much criticism of the US in the world, and we as its citizens should not criticize it.

1. Strongly agree
2. Somewhat agree
3. Somewhat disagree

4. Strongly disagree

Q39. For the most part, people who protest and demonstrate against US policy are good, upstanding, intelligent people.

1. Strongly agree
2. Somewhat agree
3. Somewhat disagree
4. Strongly disagree

Q40. If another country disagreed with an important United States policy that I knew little about, I would not necessarily support my country's position.

1. Strongly agree
2. Somewhat agree
3. Somewhat disagree
4. Strongly disagree